\documentclass[aps,twocolumn,nofootinbib,sortedaddress,reprint]{revtex4-1}
\usepackage{graphics} \usepackage{epsfig}
\usepackage{amsfonts,amsmath,amssymb,graphics,graphicx,dcolumn,bm,multirow,enumerate,verbatim,bbm,dsfont}
\usepackage{hyperref}

\begin{document}
\title{Statistical Mechanics of Competitive Resource Allocation using Agent-based Models}

\author{Anirban Chakraborti} \email{Electronic address:
  anirban@jnu.ac.in} \affiliation{Laboratoire de
  Math\'{e}matiques Appliqu\'{e}es aux Syst\`{e}mes, \'{E}cole
  Centrale Paris, 92290 Ch\^{a}tenay-Malabry, France}
  \affiliation{School of Computational and Integrative Sciences,
Jawaharlal Nehru University, New Delhi-110067, India}

\author{Damien Challet} \email{Electronic address:
  damien.challet@ecp.fr} \affiliation{Laboratoire de Math\'{e}matiques
  Appliqu\'{e}es aux Syst\`{e}mes, \'{E}cole Centrale Paris, 92290
  Ch\^{a}tenay-Malabry, France}

\author{Arnab Chatterjee} \email{Electronic address:
  arnab.chatterjee@aalto.fi} \affiliation{Department of Biomedical
  Engineering and Computational Science, Aalto University School of
  Science, P.O. Box 12200, FI-00076 AALTO, Espoo, Finland}
  \affiliation{Theoretical Condensed
  Matter Physics Division, Saha Institute of Nuclear Physics, 1/AF
  Bidhannagar, Kolkata-700064, India}

\author{Matteo Marsili} \email{Electronic address: marsili@ictp.it}
\affiliation{Abdus Salam International Centre for Theoretical Physics,
  Strada Costiera 11, 34014, Trieste, Italy}

\author{Yi-Cheng Zhang} \email{Electronic address:
  yi-cheng.zhang@unifr.ch} \affiliation{D\'{e}partement de Physique,
  Universit\'e de Fribourg, Chemin du Mus\'ee 3, 1700 Fribourg,
  Switzerland}

\author{Bikas K. Chakrabarti} \email{Electronic address:
  bikask.chakrabarti@saha.ac.in} \affiliation{Theoretical Condensed
  Matter Physics Division, Saha Institute of Nuclear Physics, 1/AF
  Bidhannagar, Kolkata-700064, India} \affiliation{Economic Research
  Unit, Indian Statistical Institute, 203 B. T. Road, Kolkata-700108,
  India}

\begin{abstract}  
Demand outstrips available resources in most situations, which gives
rise to competition, interaction and learning. In this article, we
review a broad spectrum of multi-agent models of competition (El Farol Bar problem, Minority Game, Kolkata Paise Restaurant problem, Stable marriage problem, Parking space problem and others) and the
methods used to understand them analytically. We emphasize the power
of concepts and tools from statistical mechanics to understand and
explain fully collective phenomena such as phase transitions and long
memory, and the mapping between agent heterogeneity and physical
disorder. As these methods can be applied to any
large-scale model of competitive resource allocation
made up of heterogeneous adaptive agent with non-linear
interaction, they provide a prospective unifying paradigm for many scientific disciplines.
\end{abstract}                                                                 

\maketitle \tableofcontents
\section{INTRODUCTION}
\label{sec:intro}

Most resources are in limited supply. How to allocate them is therefore of great practical importance. The variety of situations is staggering: resources may be tangible (oil, parking space, chocolates) or intangible (time, energy, bandwidth) and allocation may happen instantaneously or over a long period, and may involve a central authority or none. 

The optimal allocation of resources is a core concern of economics. The problem can be formalized as the simultaneous maximization of the utility of each member of the economy, over the set of achievable allocations. The key issue is that individuals have typically conflicting goals, as the profit of one goes to the disadvantage of the others. Therefore, the nature of the problem is conceptually different from optimization problems where a single objective function has to be maximized. 
One key insight is that markets, under some conditions, can solve efficiently the problem. This is sharply explained by Adam Smith in his famous quote:
\begin{quote}
It is not from the benevolence of the butcher, the brewer, or the baker, that we expect our dinner, but from their regard to their own interest.
\end{quote}
Markets, under some conditions, allow individuals to exchange goods for money and to reach an optimal allocation \cite{mas1995microeconomic}, where none can improve her well-being without someone else being worse off. This is called a Pareto efficient allocation.
In the exchanges, prices adjust in such a way as to reflect the value that goods have for different individuals -- codified in their marginal utilities. 

The problem with this solution are that conditions are rather restrictive: {\em i)} assumptions on the convexity of preferences or production functions have to be made. {\em ii)} The existence of markets for any commodity that individuals may be interested in, is necessary. Even if markets exist, access to them typically involves transaction costs. {\em iii)} Markets do not work for the provision of public goods, i.e., those goods whose consumption does not exclude others to draw benefit from them. {\em iv)} Markets require perfect competition where no individual has the possibility to manipulate prices. One aspect, raised by Adam Smith himself long ago in {\em The Theory of Moral Sentiments} (1759), is that market functioning requires coordination on a set of shared norms and reciprocal trust. This aspect becomes acutely evident in times of crises, when markets collapse, as we have witnessed in the 2007-08 financial crisis. 

Apart from all these issues, general equilibrium theory has provided remarkable insights on the properties of economies \cite{mas1995microeconomic}. Its strength is that it allows to relate economic behavior to the incentives that motivate the behavior of individuals, as codified in their utility functions. This opens the way to normative approaches by which policy makers may intervene in order to achieve given welfare objectives. 

Yet, the predictive power of this approach is rather limited: the mapping from the observable collective behavior to the agents' utility functions is a one-to-many. For this reason, economists have focused on specific instances where all individuals are equal, which typically reduce to problems where the economy is populated by a single {\em representative} individual. The representative agent approach has also the virtue of allowing for closed form solution, yet several of its predictions are a direct consequence of its assumptions \cite{kirman1992whom}. 
Furthermore, this approach is silent on the critical conditions that determine the stability of the system, so it provides no hints on the likelihood that markets may collapse, leading to disruption of economic activity. 

General equilibrium theory is completely silent on how the equilibrium is reached and on the conditions that allows it to be reached. This aspect has recently been addressed in the community of computer scientists, who have developed algorithms for resource allocation. There the emphasis is on decentralized heuristics for (approximate) solutions of allocation problems that are efficient in terms of computer time and can work under imperfect information. For instance, operating systems use so-called schedulers to allocate CPU and input-output resources in (almost) real time between tasks. Scalability is a primary concern, particularly in a
decentralized setting where agents need to do without global
optimization. These intrinsically non-equilibrium dynamical problems
are often solved by \textit{ad hoc} methods \cite{perkins1999adhoc},
some of them being extensions of the models that we review below
\cite{shafique2011minorityCPU}.


The collective behavior of systems of many interacting degrees of freedom has also been studied in physics. There, it has been found that the collective behavior is remarkably insensitive  to microscopic details. For example, a law of ``corresponding'' states has been derived for gases long ago (see e.g., \textcite{huang1987book}) that shows that the macroscopic behavior of real gases is well approximated by a single curve. This ``universality'' extends also to systems of heterogeneous particles, such as disordered alloys and spin glasses \cite{MPV}, thereby suggesting that a similar approach may be useful also for understanding economic phenomena. This universality allows one to focus on the simplest possible models that reproduce a particular collective behavior, which may be amenable of analytic approaches leading to a full-fledged understanding of the emergence of the collective behavior. 
Physicists have therefore applied their tools to analyze problems and model the behaviour of interacting agents in economics and sociology \cite{Chakrabarti2006,Sinha2010,Chakrabarti2013,mantegna1999,bouchaud2000,chakraborti2011a,chakraborti2011b}. \textcite{Yakovenko2009} reviewed in a colloquium, statistical models for wealth and income distributions, inspired from the  kinetic theory of gases in physics, with minimally interacting economic agents exchanging wealth. 
Another review \cite{Castellano2009}  discussed a wide range of topics from opinion formation, cultural and language dynamics,  crowd behavior, etc. where physicists studied the collective phenomena emerging from the interactions of individuals as elementary units in social structures.  Statistical mechanics of optimization problems would be well worth a review in itself (see e.g., \textcite{mezard2009information,krzakala2012compressedsensing,mezard2002k-sat}).

The present review discusses recent attempts to model and describe resource allocation problems in a population of individuals within a statistical mechanics approach. We focus on competitive resource allocation models with a fully decentralised decision process, that is, without explicit communication between the agents. We expect interaction to play a central role and to give rise to collective phenomena, such as interesting fluctuations, long memory and phase transitions. In addition the agents have a strong incentive to think and act as differently as possible in this type of competitive situation and possibly to revise their strategies \cite{Arthur}, which implies strong heterogeneity and non-equilibrium dynamics. These ingredients are very appealing to physicists, who possess the tools and concepts able to analyse and sometimes solve the dynamics of such systems, and who feel that they may contribute in a significant way to the understanding of such systems.  The usual caveat is that socio-economical agents may be orders of magnitude harder to model than, say,  electrons since they have no immutable properties and are adaptive. This, we believe, only adds to their attractiveness and, since the methods of statistical mechanics are able to solve complex models of adaptive agents, it only makes physicists' point stronger. Our aim is to provide an  account of the various mathematical methods used for this family of models and to discuss their dynamics from the perspective of physics. Due to brevity of space, we cannot provide many mathematical details and restrict the bibliography to selected topics and representative publications, but refer the reader to books and reviews.

More specifically, we consider a population of $N$  agents that try to exploit $R$ resources. Generically, we assume that $R$ denotes the number of possible choices of the agents, hence that $R\ge2$. Denote the choice of agent $i$ by $a_i\in\{1,\cdots,R\}$; his reward, or payoff, is then $u_i(\{a_{j}\})=u_i(a_i,a_{-i})$, where $a_{-i}=\{a_j\}_{j\ne i}$ contains the choices of all the agents other than agent $i$. A Nash equilibrium (NE) corresponds to a set $\{a^*_k\}$ such that $u_k(\{a^*_k\})\ge u_k(\{a_{k}\})$ $ \forall k $: it is a maximum of the payoff function and thus no agent has an incentive to deviate from his behavior \cite{FudenbergLevine}.

Section \ref{sec:MG} is devoted to the simplest case $R=2$. The agents must choose which resource to exploit, or alternatively, to exploit a resource or to abstain from it.  We shall start with the El Farol Bar Problem (EFBP) \cite{Arthur}: $N$ customers compete for $L<N$ seats at the bar. At every time step, they must choose whether to go to the bar or to stay at home. This section is then mainly devoted to the Minority Game (MG) \cite{CZ97}, which simplifies the EFBP in many respects by taking $L=N/2$. Section \ref{sec:kpr} assumes that the number of resources scales with $N$ and reviews in particular many results about the Kolkata Paise Restaurant problem (KPR), in which $R$ restaurants
have a capacity to serve one customer each, the agents trying to be alone as often as possible
\cite{kpr-physica}. Section \ref{sec:bipartite} extends the discussion to other bipartite problems, where two distinct types of agents must be matched. The Parking space problem adds space and resource heterogeneity to KPR: drivers
would like to park as close as possible from their workplace
along a linear street and must learn at what distance they are likely
to find a vacant space \cite{Hanaki2011}.   It then briefly shows the connection with the celebrated Stable marriage
problem, which assumes that $N$ men and $N$ women have their own ranking of
their potential counterparts, and studies what choosing algorithm to
apply \cite{gale1962college}. Finally, it mentions recommendation systems that try to guess the preference lists and suggest items (books, movies, etc.) to customers based on
partial information \cite{lu2012recommender}. The paper is concluded by some discussions on the approaches and on the perspective of the application of physics tools to a larger domain.

\section{Minority Games}
\label{sec:MG}
\subsection{El Farol Bar Problem}
Brian Arthur likes to listen to Irish music on Thursday evenings in El Farol bar \textcite{Arthur}. So do 99 other potential patrons. Given that there are only 60 seats in this bar, what should he do? How do his 99 competitors make up their minds, week after week? If this game is only played once, the Nash equilibrum consists in attending the concert with probability $L/N$. As a side remark, a careful customer may well count the number of seats $L$ once he is in the bar, but cannot the total number of potential patrons, which makes NE really unlikely. Real fans of Irish music do repeatedly wish to take part in the fun and use trial and error instead, an example of bounded rationality.

There are indeed many ways to be imperfect, hence, to bound rationality, while still retaining reinforcement learning abilities.   Arthur's agents are endowed with a small set of personal heuristic
attendance predictors that base their analyses on the $M$ past
attendances. They include linear predictors such as moving averages and constant ones, e.g. 42. Adaptivity consists in using the better predictors with larger likelihood. Arthur assumes that the agents trust their currently best predictors. Adaptivity also includes to discard really bad predictors and replacing them with new ones, like Darwinian evolution. 

 A seemingly remarkable result is that even with such
limited learning abilities, agents are able to self-organize and to collectively produce an
average number of bar goers (attendance) equal to $L$. It was later
realized that this was a fortunate and generic by-product of the
chosen unbiased strategy space, not the result of self-organization; other
choices are less forgiving in some circumstances \cite{CMO03}.

However, the point of Arthur remains fully valid: imperfect agents may reach a socially acceptable outcome by learning
imperfectly from random rules in a competitive setting. Ongoing competition forces the agents
to be adaptive (taken as synonym of learning) in order to outsmart
each other. There is no ideal predictor, the performance of one of them depending on the those used by all the agents.
 Arthur adds that rationality is not applicable in this setting: if everyone is rational, excluding the possibility to take random decisions
and assuming that everyone has access to the same analysis tools, 
everyone takes the same decision, which is the wrong one. Thus, negative feed-back mechanisms make
heterogeneity of beliefs or analyzes a necessary ingredient for
efficient resource allocation. In passing, heterogeneity may also emerge
  in absence of competition and negative feedback because it yields better outcomes to
  some of the players, see e.g. \textcite{matzke2011emergence}.

The very reasons of bewilderment among economists and computer
scientists who became interested in this model, were the sames ones which
triggered the interest of physicists:
\begin{itemize}
\item This model comprises interacting entities. Since they all are
  heterogeneous, interaction may also be heterogeneous.
\item Each run of the game produces a different average attendance,
  even when the average is done over an infinite number of time steps, which reminds of
  disordered systems.
\item The fact that the agents learn something may be connected in
  some way to artificial neural networks.
\item It is easy to take large values of $N$ and $L$. Intuitively,
  taking some kind of thermodynamical limit should be
  possible. This particular idea went against other fields'
    intuition at the time, see e.g. \textcite{Casti}.
\end{itemize}
The icing on the cake was the mean-field nature of this family of
models: everybody interacts with everybody else because individual
rewards depends on the choice of the whole population and rewards are
synchronized.

\subsection{From El Farol to Minority Games}

The original version of the EFBP focuses on average attendance, i.e.,
equilibrium, and has a loosely defined strategy space. Fluctuations are potentially much richer than average attendance.
  Focusing on
on them, i.e, on efficiency amounts to setting $L=N/2$ and
considering a symmetric strategy space: this is the Minority Game \cite{CZ97}. We refer the reader to the nice
  preface written by Brian Arthur in \textcite{MGbook}.
  
   It is useful to think of this model as two separate parts:
\begin{itemize}
\item The minority mechanism that is responsible for the interaction
  between the players and negative feed-back.
\item The learning scheme that determines the overall allocation
  performance.
\end{itemize}

The minority rule can be formalized as follows: each one of the $N$ agents
must select one of two options; the winners are those who choose the
least popular option, i.e., who are in the minority.  Mathematically,
if the action of agent $i$ is $a_{i}\in\{-1,+1\}$, the global action
is $A=\sum_{j=1}^{N}a_{j}\in\{-N,\cdots,+N\}$; being in the minority
is achieved if $a_{i}$ and $A$ have opposite signs; hence, the payoff
is $-a_{i}A$.

The MG is a negative sum game: the sum of all payoffs is $-\sum_i a_i
A=-A^2\le 0$, with strict inequality if $N$ is odd. This is why the
fluctuations of $A$ are a measure of global losses. Another important measure
is the asymmetry of the average outcome, i.e., the crumbs left by the
agents, measured by $H=\left<A\right>^2$, where the brackets stand for temporal average in the stationary state. When $H>0$, the outcome is
statistically predictable.

Intuitively, a stable situation is reached if no agent has an
incentive to deviate from his current behavior; this is known as a
Nash equilibrium. When all the agents have the
same expected gain, the equilibrium is called symmetric, and
asymmetric otherwise.  The Nash equilibria of the MG are discussed in
\textcite{MC00}:
\begin{enumerate}
\item A symmetric NE is obtained when all the agents toss a coin to
  choose their action; it corresponds to
  $\sigma^2=\left<A^2\right>=N$, which yields an imperfect
  allocation efficiency. Another such NE corresponds to $A=0$ (for
  $N$ even).
\item An asymmetric NE is obtained when $|A|=1$ for $N$ odd. Other
  such equilibria are reached when $n$ agents play $-1$ and $n$ play
  $+1$, while the remaining $N-2n$ play randomly. There are very
  many of them.
\end{enumerate}

\subsubsection{Re-inforcement learning and allocation efficiency}
\label{sec:mg_P1}
Simple Markovian learning schemes are well suited to familiarize
oneself with the interplay between learning and fluctuations in MGs.
Learning from past actions depends on receiving positive or negative
payoffs, reinforcing good actions and punishing bad ones.  In minority
games, as mentioned earlier, the reward to agent $i$ is $-a_{i}A$. More generally,
  the rewards may be $-a_iG(A)$ where $G$ is an odd function. The original
  game had a sign payoff, i.e. $-a_i\textrm{sign (A)}$, but linear
  payoffs are better suited to mathematics, since they are less
  discontinuous \cite{Oxf1,CMZe00}. Since learning implies playing
repeatedly, it is wise to store some information about the past in a
register usually called the score. After $t$ time steps, the score of
agent $i$ that corresponds to playing $+1$ is
\begin{equation}\label{eq:U_i_M0}
U_{i}(t+1)=-\sum_{t'=1}^{t}\frac{A(t')}{N}=U_{i}(t)-\frac{A(t)}{N}.
\end{equation}
Reinforcement learning is achieved if the probability that agent $i$
plays $+1$ increases when $U_{i}$ increases and vice-versa. It is
common to take a logit model \cite{McFadden},
\begin{equation}\label{eq:PplusP1}
P[a_{i}(t)=+1]=\frac{1+\tanh[\Gamma U_{i}(t)]}{2},
\end{equation}
where $\Gamma$ tunes the scale of reaction to a change of score; in
other words, it is a learning rate. The limit of very reactive agents
$\Gamma\to\infty$ corresponds to Arthur's prescription of playing the
best action at each time step. This defines a simple MG model
introduced and studied in
\textcite{MC00,MarsiliMinMaj,mosetti2006minority,CAM08}. If all the agent scores
start with the same initial condition, they all have the same score
evolution; hence, the dynamics of the whole system is determined by
Eq. \eqref{eq:U_i_M0} without indices $i$,  whose fixed point $U^*=0$ is unstable if $\Gamma>\Gamma_c=2$. In this case, learning takes place too rapidly; a 
finite fraction of agents reacts strongly to random fluctuations and
herds on them. This produces bimodal $A$, hence $\sigma^ 2\propto
N^2$. On the other hand, if $\Gamma<\Gamma_c$, fluctuations are of
binomial type, $\sigma^2\propto N$.  To perform better, the agents must
behave in a different way. For instance, more heterogeneity is good: the more non-uniform the initial conditions $U_i(0)$, the smaller $\sigma^2$ and the higher $\Gamma_c$ \cite{MarsiliMinMaj}.

The simple model described above does reach a symmetric equilibrium
which is not of the Nash type; one can think of it as a competitive
equilibrium. How to maximize efficiency is a recurrent theme in the
literature (see Chapter 5 of \textcite{MGbook}). A seemingly small
modification to the learning scheme described helps the agents reach
an asymmetric NE; the key point is self-impact: when evaluating the
performance of the choices $+1$ and $-1$, the agents should account for
their own impact on their payoff. More precisely, the payoff is
$-a_iA=-a_i(a_i+\sum_{j\ne i}a_j)=-1+A_{-i}$, where $A_{-i}=\sum_{j\ne
  i}a_j$: the chosen action on average yields a smaller payoff than
the other one, a generic feature of negative feedback
mechanisms. This is why \textcite{MC00} proposed to modify Eq.\ \eqref{eq:U_i_M0} into
$$ U_{i}(t+1)=U_{i}(t)-\frac{A(t)-\eta a_i(t)}{N},
$$ where $\eta$ allows agent $i$ to discount his own contribution; as
soon as $\eta>0$, the agents reach an optimal asymmetric NE ($|A|=1$).

There is a simpler way to obtain a similar result, however: laziness
(or inertia).  \textcite{Reents} assume that the agents in the
minority do not attempt to change their decision, while those in the majority do
so with fixed probability $p$. The process being Markovian, a full
analytical treatment is possible. Since there are $\frac{N+|A|}{2}$
losers, the number of agents that invert their decisions is proportional to
$pN$. Accordingly, the three regimes described above still exist
depending on $pN$.

Quite nicely, the agents never need to know the precise value of
$A(t)$, only whether they won or not; in addition, the convergence time to
$|A|=1$ is of order $\log N$ when $pN\sim 1$. This performance comes
at a cost: the agents need to choose $p$ as a function of $N$, i.e.,
they need to know the number of players, which assumes some kind of
initial synchronization or central authority.

All the above approaches do not optimize the speed of convergence to
the most efficient state. \textcite{Dhar2011} noticed that for a given
$A(t)$ the probability of switching should be such that the expected
value of $A(t+1)$ is 0, which is achieved when
\begin{equation}
\label{eq:p_Dhar}
p(t)=\frac{|A(t)|-1}{N+|A(t)|}.
\end{equation}
This dynamics holds the current record for the speed of convergence to
$|A=1|$, which scales as $O(\log\log N)$ time steps. As an illustration $\log \log 1001 \simeq 2$.
 The price to pay was of course to give even more information
to the agents: computing $p(t)$ of Eq.\ \eqref{eq:p_Dhar} requires the
knowledge of $N$ and $|A(t)|$.  This kind of dynamics was
extended further in \textcite{Biswas2012}: replacing $|A|-1$ by
$q(|A|-1)$ in Eq.\ \eqref{eq:p_Dhar} allows a dynamical phase
transition to take place at $q_c=2$: when $q>q_c$,
$\sigma^2=\frac{q-q_c}{q}N^ 2$; when $1<q<2$, $|A|$ converges to 1 in
a time proportional to $(q_c-q)^{-1}$, which duly diverges at
$q=q_c$. A similar picture emerges when each agent has his own $q_i$.

Finally, all these types of simple conditional dynamics are very
similar to those proposed in the reinforcement learning literature \cite{sutton1998reinforcement},
although nobody ever made explicit connections. This point is
discussed further in Sec. \ref{sec:mg_alg}.

\subsubsection{Original Minority Game}
\label{sec:mg_std}

The original MG follows the setup of the EFBP: it adds a layer of
strategic complexity to this setup, as the agents choose which
predictors to use rather than what actions to take. More specifically,
a predictor $a$ specifies what to do for every state of the world (which was a vector of past attendance in the EFBP).
For the sake of simplicity, we assume that the set of the states of the world
has $P$ elements. A predictor is therefore a fixed function $a$ that transforms
every $\mu\in\{1,\cdots,P\}$ into a choice $a^\mu\in\{-1,+1\}$, which is nothing else than a vector of binary choices
that the literature on the
MG prefers to call strategies. There are $2^{P}$ of them. Since $P$ does not depend on $N$ in any way this ensures that  that one can define a proper
thermodynamic limit, in contrast with the EFBP. In addition, one already can predict that fluctuations
are likely to be large if there are many more agents than available
strategies: some strategies will be used by a finite fraction of
agents, leading to identical behavior, or herding.

 In the original MG, $\mu$ is a number corresponding to the binary
 encoding of the last $M$ past winning choices, $M$ being the history
 length, hence, $P=2^M$. Note that simple MGs discussed in Sec.
 \ref{sec:mg_P1} will be referred to as $P=1$ henceforth, since $M=0$.

Adaptivity consists of being able change one's behavior and is highly
desirable in a competitive setting. In the original MG, the agents need
therefore at least two strategies to be adaptive; for the sake of
simplicity, we shall only consider here agents with two strategies
$a_{i,s}$ where $s$ can take two values; it is advantageous to denote
them $s_{i}\in\{-1,+1\}$. The case $S>2$ is investigated in e.g.
\textcite{Savit2,MCZe00,CoolenS>2,AdemarS>2}.

In addition, the agents use reinforcement learning on the strategies,
not on the bare actions. One thus attributes a score $U_{i,s}$ to each
strategy $a_{i,s}$ that evolves according to
\begin{equation}\label{eq:U_i_M>0}
U_{i,s}(t+1)=U_{i,s}(t)-a_{i,s}^{\mu(t)}(t)\frac{A(t)}{N},
\end{equation}
where $A(t)=\sum_{i=1}^N a_{i,s_i(t)}^{\mu(t)}$ and $s_i(t)$ denotes
the strategy played by agent $i$ at time $t$. Using also a logit
model, as in Eq.~\eqref{eq:U_i_M0}, one writes for $S=2$ 
\begin{align}\nonumber
P[s_{i}(t)=+1]&=\frac{e^{\Gamma U_{i,+}(t)}}{e^{\Gamma
    U_{i,+}(t)}+e^{\Gamma U_{i,-}(t)}}\\ &=\frac{1+\tanh[\Gamma
    (U_{i,+}(t)-U_{i,-}(t))/2]}{2}\label{eq:PplusP>1}.
\end{align}

The original MG follows Arthur's `use-the-best' prescription, which
corresponds to $\Gamma\to\infty$, while finite $\Gamma$ was introduced
by \textcite{Oxf1} as an inverse temperature. Extrapolating the
results for $P=1$ in Sec. \ref{sec:mg_P1}, one expects some herding
for $\Gamma$ larger than some value provided that $N$ is ``large
enough''.

Equation \eqref{eq:PplusP>1} shows that the choice of a strategy only
depends on the difference of scores for $S=2$. It is therefore useful to
introduce $Y_i=\Gamma(U_{i,+}-U_{i,-})/2$ and
$\xi_{i}=(a_{i,+}-a_{i,-})/2$; Eq.\ \eqref{eq:U_i_M>0} then becomes
\begin{equation}\label{eq:q_i(t)}
Y_{i}(t+1)=Y_{i}(t)-\Gamma\xi_{i}^{\mu(t)}(t)\frac{A(t)}{N}.
\end{equation}
If one denotes $\omega_{i}=(a_{i,+}+a_{i,-})/2$, the individual action
can be written
\begin{equation}\label{eq:a_omega_xi_si}
a_i(t)=\omega_i^{\mu(t)}+\xi_i^{\mu(t)}s_i,
\end{equation}
 and thus  $A(t)=\sum_{i=1}^N\omega_i^{\mu(t)}+\sum_i\xi_i^{\mu(t)}s_i(t)$,
 which strongly suggests to introduce
 $\Omega^\mu=\sum_i\omega_i^{\mu(t)}$; finally
\begin{equation}\label{eq:A_Om_xi}
A(t)=\Omega^{\mu(t)}+\sum_{i=1}^N\xi_i^{\mu(t)}s_i(t).
\end{equation}

\begin{figure}
\includegraphics*[width=8.5cm]{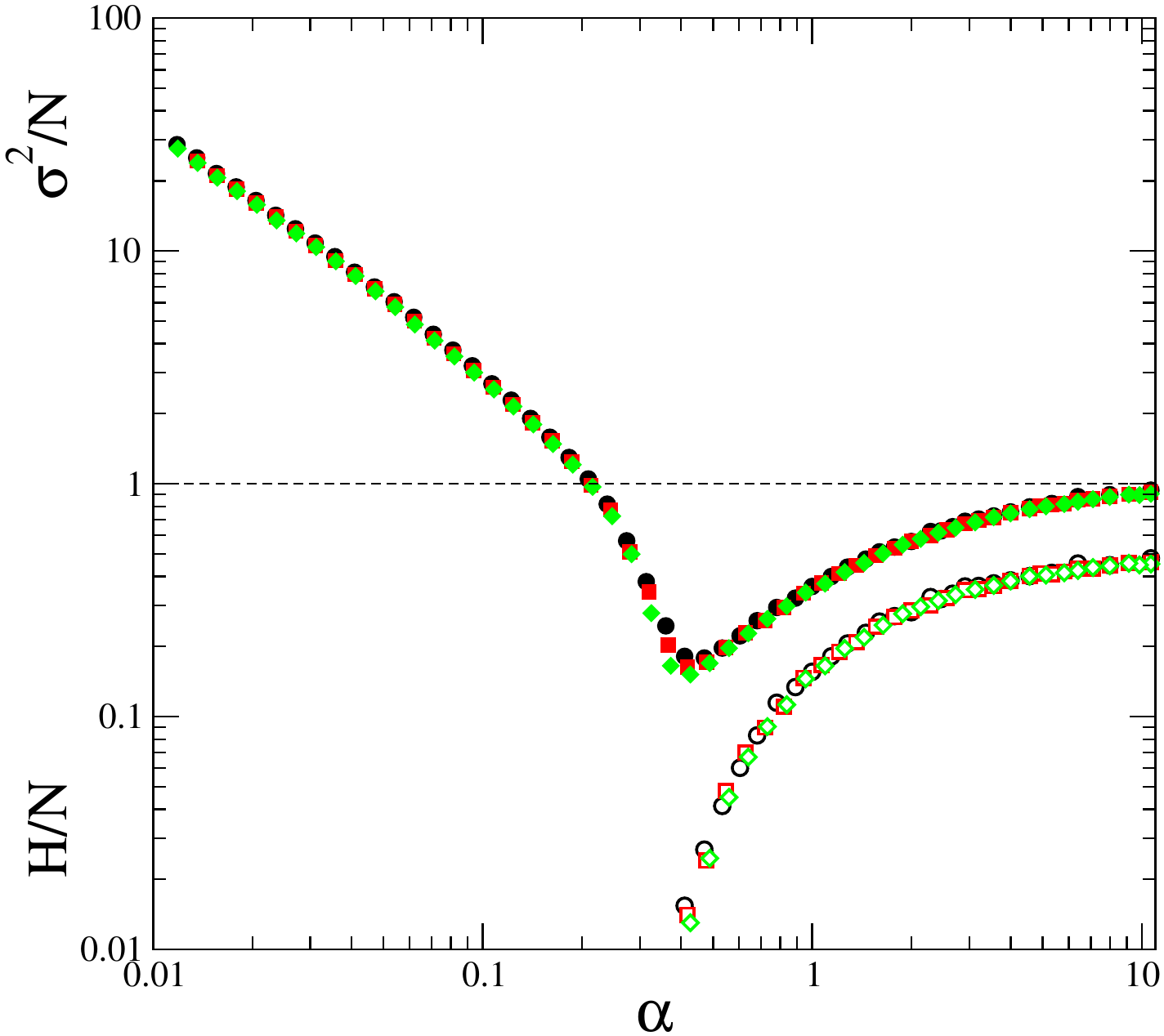}
\caption{Scaled fluctuations $\sigma^2/N$ (filled symbols) and scaled
  predictability $H/N$ (open symbols) as a function of $\alpha=P/N$ for  $P=32$, 64 and 128 (circles, squares and diamonds,
  respectively); averages over 200 samples taken over $200P$ time steps after a waiting time of $200P$ time steps}
\label{fig:s2HRS} 
\end{figure}
\textcite{Savit} showed that the control
parameter of the MG is $\alpha=P/N$. In other words, properly rescaled
macroscopic measurables are invariant at fixed $\alpha$ for instance
when both $P$ and $N$ are doubled; this opens the way to systematic
studies and to taking the thermodynamic limit. When performing
numerical simulations, too many papers overlook the need to account
first for the transient dynamics that leads to the stationary state,
and many more that this model has an intrinsic timescale,
$P$. Intuitively, this is because the agents need to explore all
answers from both their strategies in order to figure out which one is
better than the other one. As a rule of thumb, one is advised to wait
for $2000P/\Gamma$ iterations and to take averages over the next $2000P/\Gamma$, or $200P$ each for $\Gamma=\infty$
iterations, although this rough estimate is probably too small near a critical point
(see Sec. \ref{sec:signal-noise}).

Figure \ref{fig:s2HRS} reports scaled fluctuations $\sigma^2/N$ for
various values of $P$ as function of $\alpha$. The generic features of
this kind of plot are the following:
\begin{enumerate}
\item The collapse is indeed excellent, up to finite size effects.
\item In the limit $\alpha\to\infty$, $\sigma^2\to N$, which
  corresponds to random strategy choices; since the latter are
  initially attributed randomly, the resulting actions are also random.
\item In the limit $\alpha\to0$, $\sigma^2\propto N^2$, which means
  that a finite fraction of agents is synchronized, i.e., herds.
\item There is a clear minimum of $\sigma^2/N$ at $\alpha\simeq0.4$
  whose precise location depends on $N$ (see also
  Fig. \ref{fig:signal-noise}).
\end{enumerate}

\textcite{Savit} also note that the average sign of $A$ conditional to
a given $\mu$ is zero for $\alpha<\alpha_c$ and systematically
different from zero for $\alpha>\alpha_c$, which means that there is some predictability in the asymmetric phase.
 Denoting the average of $A$
conditional to $\mu$ as $\left<A|\mu\right>$, one defines a smoother conditional predictability
\begin{equation}\label{eq:H_def}
H=\frac{1}{P}\sum_{\mu=1}^P\left<A|\mu\right>^2.
\end{equation}
 Figure \ref{fig:s2HRS}
reports the behavior of $H/N$ as a function of $\alpha$: there is a
transition at $\alpha_c$ where $H$ is cancelled. In addition $H/N$ behaves in a smooth way close to the transition $\alpha-\alpha_c\ll1$. Since $H$ is a
measure of asymmetry, this behavior is in fact tell-tale of a  second-order phase
transition with broken symmetry. Accordingly, the two phases are known
as {\em symmetric }($H=0$) and {\em asymmetric} ($H>0$), or
(un-)predictable.
\begin{figure}
\includegraphics*[width=8.5cm]{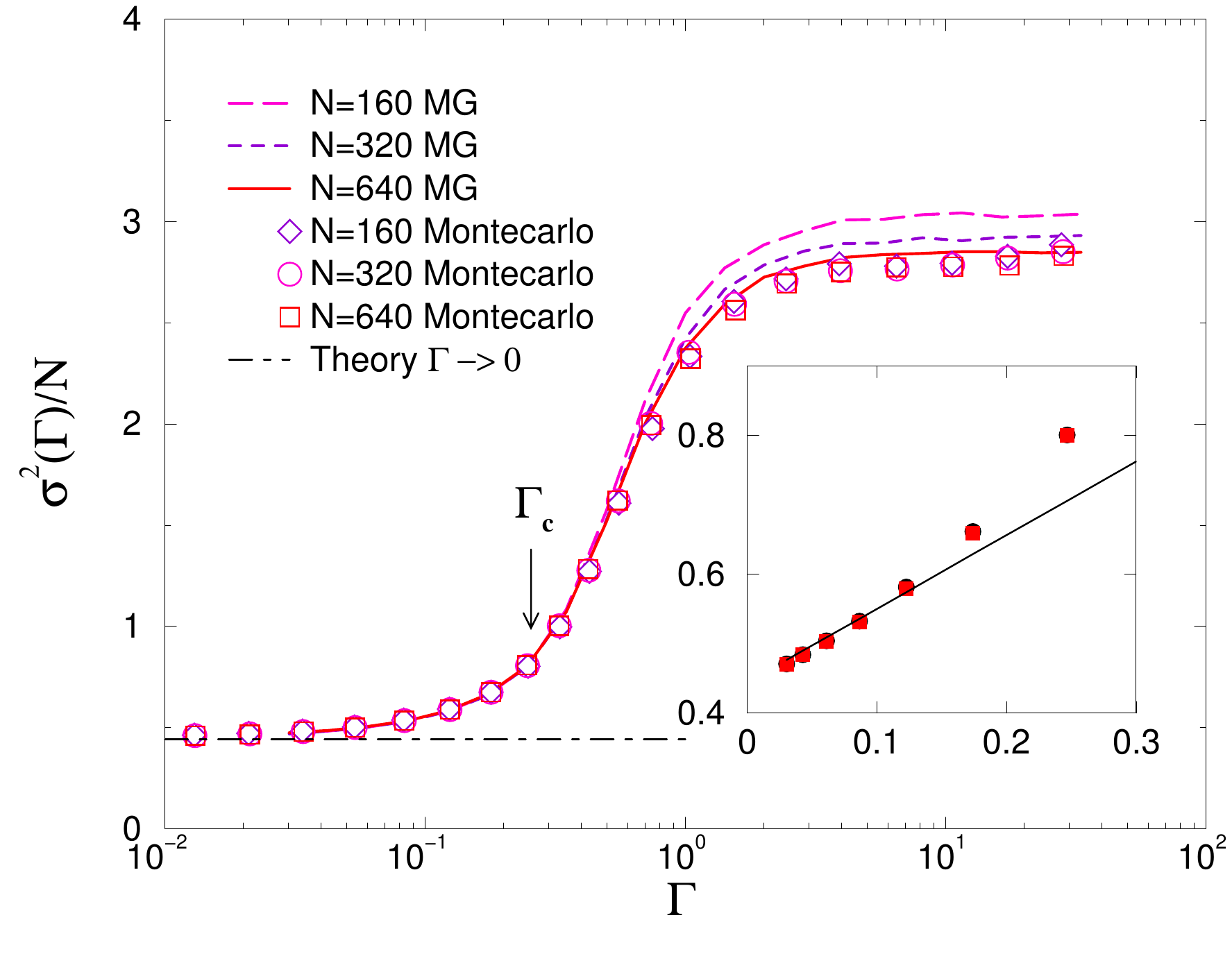}
\caption{\label{fig:s2gamma} Scaled fluctuations $\sigma^2/N$ as a
  function of $\Gamma$ at $\alpha=0.1$. Inset: $\sigma^2/N$ vs $\Gamma$; continuous line: prediction from Eq. \eqref{eq:s2_gamma}. Averages of 100 samples.
  From \textcite{MC01}.}
\end{figure}

Provided that the agents are given look-up tables, the presence of a
phase transition is very robust with respect to changes in the choice
of strategies and various sources of noise in the decision-making process.
 \textcite{galla2008transition} show that MGs with look-up
tables $a^\mu$ undergo this kind of phase transition as long as a
finite fraction of the agents behaves as those of the original MG. The
stationary state does not depend on the value of $\Gamma$
 in the asymmetric phase, nor does the location of the phase transition. This is remarkable, as this
parameter was introduced as an inverse temperature, but it is not able to cause a phase transition. In fact,
it is rather the time scale over which the agents average the
fluctuations of their $Y_i$s. In the symmetric phase, however, as
shown in Fig.\ \ref{fig:s2gamma}, the fluctuations decrease when
$\Gamma$ decreases \cite{DosiExp,Oxf1}; this is because adding noise
to the decision process breaks the herding tendency of the agents by
decreasing the sensitivity of the agents to fluctuations of their
payoffs, exactly as in the $P=1$ case. Even more, one can show (see Sec.~\ref{sec:signal-noise})
that the dynamics becomes deterministic when $\Gamma\to0$. In this
limit $\sigma^2/N<1$ for all values of $\alpha$ \cite{MC01}.

Initial conditions, i.e., initial score valuations, have an influence
only the stationary state of the symmetric phase, i.e., on the
emergence of large fluctuations \cite{dhulst2000strategy,Moro1}. The
insights of the $P=1$ case are still valid \cite{MarsiliMinMaj}: large
fluctuations are killed by sufficiently biased initial
conditions. This point will be discussed again in Sec.
\ref{sec:mg_maths}.

\subsection{Mathematical approaches}
Statistical mechanics has been applied successfully to two-player
games that have a large number of possible choices
\cite{galla2013complex,berg1998matrix,berg2000statistical}. The MG
case is exactly the opposite: two choices, but very many players, with
proportionally many states of the world.

\label{sec:mg_maths}
\subsubsection{Algebra: why is there a critical point?}
\label{sec:mg_alg}
Before understanding why $H=0$ for $\alpha<\alpha_c$, it is wise to
investigate why $H=0$ is possible at all \cite{MC01}. From Eq.~\eqref{eq:H_def},
setting $H=0$ requires all conditional averages to be zero, i.e.,
$\left<A|\mu\right>=0$, i.e., from Eq.\ \eqref{eq:A_Om_xi},
\begin{equation}\label{eq:Amu_eq_Om}
\sum_i\xi_i^\mu \left<s_i\right>=-\Omega^\mu.
\end{equation}
It helps thinking of $\left<s_i\right>$ as a continuous variable
$m_i\in[-1,1]$: achieving $H=0$ requires to solve a system of $P$
linear equations of $N$ variables.

This set of equations yields surprisingly many insights on the
stationary state of minority game-like models:

\begin{enumerate}
\item The fact that $\alpha_c\simeq 0.4 < 1$ means that one needs more that $P$ variables to
  solve this set of equations; this is because the $m_i$s are bounded.
\item The control parameter is the ratio between the number of equations and the number 
of variables, $P/N$, and not $2^P/N$, i.e., the total
  number of possible strategies per agent.
\item $m_i=0$ $\forall i$ is always a solution if all
  $\Omega^\mu=0$. In other words, if all the agents have two opposite
  strategies, one predicts that $\sigma^2/N=1$ for $\alpha>\alpha_c$,
  and that $\alpha_c=1$; the exact solution confirms this intuition
  \cite{MMM}. What happens for $\alpha<\alpha_c$ is similar whatever
  the distribution of $\Omega^\mu$: the degrees of freedom not needed
  to cancel $H$ allow the agents to herd and synchronize; as a
  consequence, $\sigma^2\propto N^2$.
\item Since $m_i$s are bounded, some agents have $|m_i|=1$ when the
  equations are not satisfied: those agents always play the same
  strategy. They are fittingly called \emph{frozen} \cite{CM99}. Once
  frozen, the contribution of an agent is fixed, hence can be
  incorporated into the fields $\Omega^\mu$. Accordingly, the number
  of degrees of freedom decreases; denoting the fraction of frozen
  agents by $\phi$, the remaining number of degrees of freedom is
  $(1-\phi)N$. At $\alpha_c$, the number of degrees of freedom  must equate $P$,
  i.e., $\alpha_c=1-\phi_c$ \cite{MCZe00,MC01}. This intuition is confirmed by the exact
  solution of the model (see Sec. \ref{sec:replica}).
\item This set of equations specifies which subspace is spanned by the
  dynamic variables in the stationary state
  \cite{MC01}. \textcite{GallaClubbing} noted that as long as the
  dynamics is of the form $Y_i(t+1)=Y_i(t)-\xi_i^\mu F(A(t))$ for some
  function $F$, a similar set of equations is solved by the dynamics;
  however, if $A$ acquires dependence on $i$, or if $Y(t)$ is multiplied by a discount factor, or if a
  constant is added to the payoff, no such set of equations holds and
  no phase transition is found.
\end{enumerate}

\subsubsection{Continuous time}
\label{sec:mg_continuoustime}
\textcite{MC01} derive the continuous-time limit of
Eq.\ \eqref{eq:q_i(t)}. The key idea is to average the payoffs to
agents in a time window of length proportional to the intrinsic time 
scale of the system, $P/\Gamma$, thus, to define the continuous time
$\tau=\Gamma t/N$.\footnote{\textcite{Moro1} derives an effective
  dynamics without taking timescales into account, which reproduces the global behavior of $\sigma^2/N$ approximately.} In the
thermodynamic limit, at fixed $\alpha$, $\tau$ becomes
continuous. Finally, setting $y_i(\tau)=\lim_{N,P\to\infty} Y_i(t)$,
one finds
\begin{align}\label{eq:dy}
\frac{dy}{d\tau}&=-\overline{\xi_{i}^\mu\left<A(\tau)|\mu\right>}_y+\zeta_i\\ &=h_i+
\sum_j J_{i,j}\tanh(y_j)+\zeta_i,
\end{align}
where the average $\left<.\right>_y$ is over the distribution of the
$m_i$s at time $\tau$, i.e., depends on the $y_i$s at time $\tau$,
$h_i=\frac{1}{P}\sum_\mu \xi^\mu\Omega^\mu=\overline{\xi^\mu\Omega^\mu}$ and
$J_{i,j}=\overline{\xi_i^\mu\xi_j^\mu}$ and the noise term is
\begin{align}
\left<\zeta_i(\tau)\right>&=0\\ \left<\zeta_i(\tau)\zeta_j(\tau')\right>&=\frac{\Gamma}{N}\overline{\xi_i^\mu\xi_j^\mu\left<A^2|\mu\right>_y\delta(\tau-\tau')}.
\end{align}
This shows that  the dynamics becomes deterministic when $\Gamma=0$,.

The autocorrelation of the noise term does not vanish in the
thermodynamic limit. Even more, it is proportional to the
instantaneous fluctuations, which makes sense: this reflects the
uncertainty faced by the agents, which is precisely $\sigma^2$. This
is in fact a powerful feedback loop and is responsible for the
build-up of fluctuations near the critical point. Deep in the
asymmetric phase, this feedback is negligible, thus
$\overline{\xi_i\xi_j\left<A^2\right>_y}\simeq
J_{i,j}\,\overline{\left<A^2\right>_y}\simeq J_{i,j} \sigma^2$ is a good approximation, which consists of equating the instantaneous volatility to the stationary
volatility. This is fact is a very good approximation over the whole range of $\alpha$. 
The noise autocorrelation takes then a form familiar to
physicists,
\begin{equation}
\left<\zeta_i(\tau)\zeta_j(\tau')\right>\simeq
2TJ_{i,j}\delta(\tau'-\tau), \textrm{ with
}T=\frac{\Gamma\sigma^2}{2N}.
\end{equation}
This effective theory is self-consistent: $T$ depends on $\sigma$,
which depends on $y_i$, which depends on $T$. The probability
distribution function $P(\{y_i\})$ in the stationary state is given in
\textcite{MC01}.

The derivation of continuous-time dynamics makes it possible to apply
results from the theory of stochastic differential equations. Using
Veretennikov's theorem \cite{veretennikov2000polynomial},
\textcite{ortisi2008polynomial} derives an upper bound to the speed of
convergence to the stationary state which expectedly scales as
$N/\Gamma$ for $\alpha>\alpha_c$.

\subsubsection{Signal-to-noise ratio, finite size effects and large fluctuations}
\label{sec:signal-noise}
\begin{figure}
\includegraphics*[width=8.5cm]{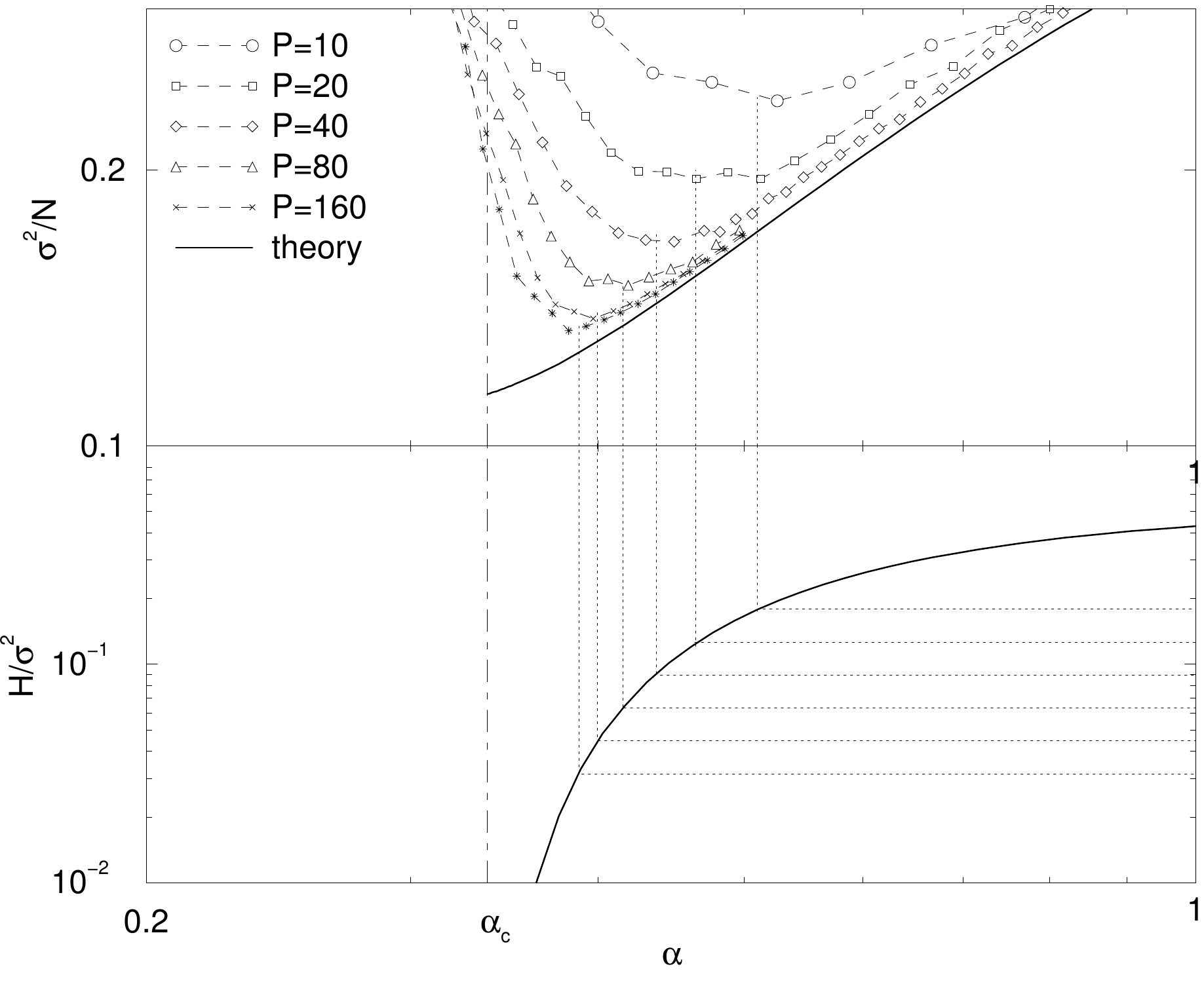}
\caption{\label{fig:signal-noise} Top panel: scaled fluctuations
  $\sigma^2/N$ as a function of $\alpha$ for increasing $P$; bottom
  panel: signal-to-noise ratio $H/\sigma^2$ from the exact solution
  together with continuous lines at $K/\sqrt{P}$; $K\simeq 0.39$. From
  \textcite{MGbook}.}
\end{figure}

Figure \ref{fig:signal-noise} shows the existence of finite-size
effects near $\alpha_c$. In particular, the larger the system size,
the smaller the minimum value of $\sigma^2/N$ and the smaller the
location of its minimum. To understand why this happens, one has to
take the point of view of the agents, i.e., of their perception of the
world, which is nothing else than Eq.\ \eqref{eq:dy}. The fluctuations
of the score of agents $i$ and $j$ become correlated via their noise
terms if the strength of the latter becomes comparable to that of their payoffs,
i.e., when $K\sqrt{\Gamma J_{i,j}\sigma^2/N}=\sqrt{H/P}$, where $K$ is
a proportionality factor. Since $J_{i,j}\propto P^ {-1/2}$, this
condition becomes, by incorporating $\sqrt{\Gamma}$ into $K$,
\begin{equation}
\frac{H}{\sigma^2}=\frac{K}{\sqrt{P}}.\label{eq:sign-noise-MG}
\end{equation}
$H$ and $\sigma^2$ are known from the exact solution for infinite
systems (see Sec.~\ref{sec:replica}). The above intuition is
confirmed by numerical simulations and the exact
solution (Fig.\ \ref{fig:signal-noise}): one sees that the intersection between $K/\sqrt{P}$ and the ratio
$H/\sigma^2$ given by the exact solution predicts the point at which
$\sigma^2/N$ deviates significantly from the exact solution, defined
as the locus of its minimum. Since $H\propto (\alpha-\alpha_c)^2$ for
$\alpha-\alpha_c\ll1$ (see Sec.
\ref{sec:replica}), the size of this region scales as
$N^{-1/4}$. Similar transitions are found in all MGs in which the
noise may acquire a sufficient strength, in particular in market-like
grand-canonical games (see Sec. \ref{sec:mg_markets}); the
procedure to find them is the same: derive continuous time equations,
compute the inter-agent noise correlation strength, and match it with
the drift term. This transition is ubiquitous: it happens
in any model underlaid by a minority mechanism when the agents do not
account for their impact.\footnote{A recent generic result about optimal learning and emergence of anomalous noise when nothing much remains to be learnt leads to comparable results \cite{patzelt2011criticality}.}

\subsubsection{Reduced set of strategies}

A naive argument suggests that herding should occur when a finite
fraction of agents adopt the same strategies, hence that $\alpha=2^
P/N$. The problem lies in the definition of ness: the fraction of
different predictions between strategies $a$ and $b$ is the Hamming
distance 
\begin{equation}
d(a,b)=\frac{1+\frac{1}{P}\sum_\mu a^\mu b^\mu}{2}.
\end{equation} 
For large $P$, two strategies do not differ by much if they differ by only
one of their predictions. \textcite{ZENews} defines three levels of
sameness: either same ($a=b$), opposite ($a=-b$), or uncorrelated
($d(a,b)=1/2$). Starting from an arbitrary strategy $a$, there are
exactly $2\times 2^M$ strategies that are either same, opposite, or
uncorrelated with $a$ and with each other \cite{CZ98}; this is called
the reduced strategy set (RSS). Forcing the agents to draw their
strategies from this set yields very similar
$\frac{\sigma^2}{N}(\alpha)$ \cite{CZ98}. Now, since
$\sigma^2=N+\sum_{i\ne j}\left<a_ia_j\right>$, the RSS allows to
decouple the correlation term into the contributions of uncorrelated,
correlated and anti-correlated agents; the latter two are known as
herds and anti-herds, or crowds and anti-crowds. Thus, the final value
of the fluctuations can be seen as the result of competition between
herding and anti-herding. This yields several types of analytical
approximations to the fluctuations that explain the global shape of
$\sigma^2/N$ as a function of $\alpha$. More generally, as it reduces
much the dimension of the strategy space, this approach simplifies the
dynamics of the model and allows one to study it in minute details; it
has been has been applied to a variety of extensions
\cite{JohnsonCrowds,JohnsonDeterministic,JohnsonCrowdsTheory,choe2004errordriventransition}.
In addition, when the agents only remember the last $T$ payoffs, 
the whole dynamics is Markovian of order $T$; simple analytical formulations give many 
insights about the origin of large fluctuations \cite{JohnsonHorizon,satinover2008cycles}.

\subsubsection{The road to statistical mechanics}

A great simplification comes from the fact
that the global shapes of $\sigma^2/N(\alpha)$ and $H/N(\alpha)$ are
mostly unchanged if one replaces the bit-string dynamics of $\mu$s
with random $\mu$ drawn uniformly with equal probability  \cite{Cavagna}.\footnote{It
  was initially believed that only the frequency distribution with
  which each $\mu$ appears had an influence on $H$ and $\sigma^2$
  \cite{CM00}, hence that $\sigma^2$ did not depend on the nature of
  histories in the symmetric phase. Later work showed that finite-size
  effects where responsible for this apparent independence and that
  periodic, random or real histories lead to different $\sigma^2/N$
  (see e.g. \textcite{hung2007effective}). Finally, the exact
  dynamical solution of MGs with real histories was derived in a
  rigorous way in \textcite{CoolenRealHistories}.}

In short, two methods are known to produce exact results: the replica
trick and generating functionals {\em \`a la}
\textcite{DeDominicis}. The replica trick is simpler but less
rigorous; in addition it requires to determine what quantity the
dynamics minimizes, which is both an advantage as this quantity
reveals great insights about the global dynamics and a curse as there
may be no discernible minimized quantity, precluding the use of this
method. Generating functionals consist of rigorous {\em ab initio} calculus
and does not require the knowledge of the minimized quantity, which is
both regrettable and a great advantage (invert the above statements about the replica calculus).
The following account aims at giving the spirit of these methods
and what to expect from them, i.e., their main results, their level of
complexity and their limitations.

For lack of space, we can only give the principles of the methods in
question. Detailed calculus is found in
\textcite{demartino2006statistical}, who deal with statistical
mechanics applied to multi-agent models of socio-economic systems,
\textcite{galla2006anomalous} who review anomalous dynamics in
multi-agent models of financial markets, and the two books on the MG
\cite{MGbook,CoolenBook}.

\subsubsection{Replica}
\label{sec:replica}
The drift term in Eq.\ \eqref{eq:dy} contains the key to determine the
quantity minimized by the dynamics: one can write
$2\overline{\xi_{i}^\mu\left<A\right>}_y=\frac{\partial H}{\partial m_i}$; therefore,
the predictability $H$ is akin to a potential. When $\Gamma=0$, the
dynamics is deterministic and $H$ is a Lyapunov function of the system
and is minimized; when $\Gamma>0$, $H$ still tends to its minimum. A
similar line of reasoning applies to non-linear payoffs $-a_ig(A)$ and
yields more intricate expressions \cite{MC01}.

Let us focus on linear payoffs. Given its mathematical definition, $H$
possesses a unique minimum as long as $H>0$, which determines the
properties of the system in the stationary state. Regarding $H$ as a
cost function, i.e., an energy, suggests to use a partition function
$Z=\textrm{Tr}_{\{m_i\}}e^{-\beta H}$, which yields the minimum of $H$
at zero temperature
\begin{equation}
\min_{\{m_i\}} H=-\lim_{\beta\to \infty}\frac{1}{\beta}\log Z.
\end{equation}
This only holds for a given realization of the game, i.e., for a given
set of agents, which is equivalent to fixed (quenched) disorder in the language of
physics. Averaging over all possible strategy attributions is easy in
principle: one computes
$\left<H\right>_{\{a_i\}}=-\lim_{\beta\to\infty}\frac{1}{\beta}\left<\log
Z\right>_{\{a_i\}}$. Averages of logarithms are devilishly hard to compute,
but the identity $\log Z = \lim_{n\to0}\frac{Z^n-1}{n}$ leaves some
hope: one is left with computing $\left<Z^n\right>_{\{a_i\}}$, which must be
interpreted as $n$ replicas of the same game running simultaneously, each
with its own set of variables. The limit
$n\to0$ is to not to be taken as annihilation, but as analytical
continuation.

Finally, one takes the thermodynamic limit, i.e., $P$, $N\to\infty$ at
fixed $P/N=\alpha$. In this limit, the fluctuations of global
quantities induced by different strategy allocations vanish: the system
is called {\em self-averaging}. In passing, this implies that numerical
  simulations require less samples as the size of the system decreases
  in order to achieve similar error bars.

As usual, one loves exponentials of linear terms when computing
partition functions.  $H$ is a sum of squared terms that are transformed 
into linear terms averaged over  Gaussian auxiliary
variables. This finally yields
\begin{equation}
H_0=\lim_{N\to\infty}\frac{H}{N}=\frac{1+Q_0}{2(1+\chi)^2},
\end{equation}
where $Q_0=\lim_{N\to\infty}\frac{1}{N}\sum_{i=1}^Nm_i^2$ measures
strategy-use `polarization' and $\chi$ is the integrated `response' to a
small perturbation. These two quantities are defined as

\begin{align}
Q_0&=1-\sqrt{\frac{2}{\pi}}\frac{e^-\zeta^2}{\zeta}-\left(1-\frac{1}{\zeta^2}\right)\textrm{erf}\left(\frac{\zeta}{\sqrt{2}}\right),\\ \chi&=\frac{\textrm{erf}\left(\frac{\zeta}{\sqrt{2}}\right)}{\alpha-\textrm{erf}\left(\frac{\zeta}{\sqrt{2}}\right)},
\end{align}
where $\zeta$ is determined for $\alpha>\alpha_c$ by
\begin{equation}
\alpha=[1+Q_0(\zeta)]\zeta^2.
\end{equation}
Hence, $\zeta$ is a function of $\alpha$ and determines all the above quantities. Since $\zeta>0$, this equation is easily solved recursively by writing
$\zeta_{n+1}=\sqrt{\alpha/[1+Q_0(\zeta_n)]}$, with $\zeta_0=0.5$.

$H_0=0$ is only possible when $\chi=\infty$, i.e., when the response
function diverges. This happens at the phase transition, which implies
that
\begin{align}
\alpha_c&=\textrm{erf}\left(\frac{\zeta_c}{\sqrt{2}}\right)=0.3374\dots
\end{align}
and that $H\propto (\alpha-\alpha_c)^2$ near the critical point in the
asymmetric phase.

The full distribution of $m_i$ is given by
\begin{equation}
P(m)=\frac{\phi}{2}\delta(m-1)+\frac{\phi}{2}\delta(m+1)+\frac{\zeta}{\sqrt{2\pi}}e^{-(\zeta
  m)^2/2},
\end{equation}
where $\phi=\textrm{erfc}\left(\frac{\zeta}{\sqrt{2}}\right)$ is the
fraction of frozen agents. Incidentally, this confirms that
$\phi=1-\alpha_c$ at the critical point, as guessed in Sec.~\ref{sec:mg_alg}.

The fluctuations $\sigma^2_0=\lim_{N\to\infty}\sigma^2/N$ do not
depend on initial conditions for $\alpha>\alpha_c$ and are given by
\begin{equation}\label{eq:s2RS}
\sigma^2_0=H_0+\frac{1}{2}(1-Q_0).
\end{equation}
The more rigorous generating functionals discussed below reproduce all
the above equations and bring many more insights on the
dynamics. They also show in what limit Eq.\ \eqref{eq:s2RS} is
correct \cite{CoolenOnline}.

The case $\alpha<\alpha_c$ is more complex. The good news is that the above equations
equation are still valid when $\Gamma=0$. By introducing Gaussian
fluctuations around the stationary values of $m_i$, \textcite{MC01}
give the first-order expansion
\begin{equation}
\label{eq:s2_gamma}
\sigma_0^
2=\frac{1-Q_0}{2}\left[1+\frac{1-Q_0+\alpha(1-3Q_0)}{4\alpha}\Gamma+O(\Gamma^2)\right]
\end{equation}
whose validity can be checked in the inset of
Fig.\ \ref{fig:s2gamma}. Furthermore, \textcite{MC00} derive
$\Gamma_c(\alpha)$ above which $\sigma^2$ becomes of order $N^2$. As
shown in Fig.\ \ref{fig:s2gamma}, $\sigma^2/N$ reaches its large value
plateau for a finite value of $\Gamma$; as a consequence, the limit
$\Gamma\to\infty$ can be interpreted as equivalent to {\em large
  enough} $\Gamma$.

Replica calculus has been extended to account for biased initial conditions in the
symmetric phase in an indirect way; for example, the limit of infinite
bias yields $\sigma^2_0\propto \alpha$ for small $\alpha$ \cite{MC01}.

Finally, replica calculus for games with $S>2$ is found in
\textcite{MCZe00}. \textcite{CM00} take into account the diversity of
frequency of the $\mu$ in games with real histories.  Replica calculus
can also be applied to the extensions discussed in Sec.
\ref{sec:mg_ext} that have a discernible cost function.

\subsubsection{Generating functionals}

Generating functionals keep the full complexity of the dynamics of the
model in an elegant way \cite{DeDominicis}. The reasoning is as
follows: the state of a MG at time $t$ is given by the vector of score
differences $\mathbf{Y}=\{Y_{i}\}$; one is thus interested in $P_t(\mathbf{Y})$ and
in its evolution, written schematically as
\begin{align}
P_{t+1}(\mathbf{Y'})&=\int
d\mathbf{y}P(\mathbf{Y})W_t(Y'|Y)\\ W_t(\mathbf{Y'}|\mathbf{Y})&={\prod_i\delta[Y_i'-(Y_i-\Gamma\xi_i
    ^ {\mu(t)}A(t)/N)]},
\label{eq:testYY}
\end{align}
where one recognizes Eq.\ \eqref{eq:q_i(t)} in
Eq.\ \eqref{eq:testYY}. This suggests a way to describe all the
possible paths of the dynamics: the generating functional of the
dynamics is
\begin{equation}
Z[\mathbf{\psi}]=\left<e^{i\sum_t\sum_i\psi_{i,t,}s_{i,t}}\right>_\mathrm{paths,
  disorder},
\end{equation}
from which one can extract meaningful quantities by taking derivatives
of $Z$ with respect to the auxiliary variables, for instance $\partial
Z/\partial \psi_{i,t}=\left<s_i(t)\right>$. An important point is that
the really hard work resides in taking the average over all possible
paths. What is multiplied by $\psi_{i,t}$ can be chosen at will
depending on what kind of information one wishes to extract from
$Z$. In addition, it is very useful to add a perturbation to the
dynamical equations, so that
$W_t(\mathbf{Y'}|\mathbf{Y})={\prod_i\delta[Y_i'-(Y_i-\xi_i^{\mu(t)}
    A(t)/N)+\theta_{i,t}]}$: taking derivatives of $Z$ with
respect to $\theta_{i,t}$ yields response functions of the system.

Nothing prevents in principle to include the dynamics of $\mu$ and
thus solve the full original MGs. Given the length of the calculus, it
is worth trying to simplify further the dynamics of $\mu$ and $s_i$ by
assuming that the agents update $s_i(t)$ every $P$ time steps, and
that the $\mu$'s appear exactly once during this interval. This is
called the batch minority game \cite{Moro1}, while the update of the
$\{s_i\}$'s at each time step is referred to as on-line. Crucially,
once again, the global shape of $\sigma^2$, $H$ and $\phi$ are left
intact. Batch games lead to a simpler $W$, which reads now
$W_t(\mathbf{Y'}|\mathbf{Y})=\prod_i\delta[Y_i'-(Y_i-\overline{\xi_i
    A/N})+\theta_{i,t}]$.

The calculus is long: after putting the last Dirac and Heaviside
function in an exponential form, and removing all the non-linear terms
of the argument of the exponential with auxiliary variables, one
performs the average of the disorder (i.e., strategy assignment) and
then take the thermodynamic limit. One is then usually rewarded by the
exact effective agent dynamics
\begin{equation}\label{eq:eff_dyn}
Y(t+1)=Y(t)+\theta-\alpha\sum_{t'\le
  t}(\mathbbm{1}+G)_{tt'}^{-1}\textrm{sgn}\,Y(t')+\sqrt{\alpha}\eta(t),
\end{equation}
where $G_{tt'}=\lim_{N\to\infty}\frac{1}{N}\sum_i\frac{\partial
}{\partial \theta_{i,t}}\left<s_i(t')\right>_\mathrm{paths, disorder}$
is the average response function of the spins encoding strategy choice
at time $t$ to a perturbation applied at an earlier time $t'$ and
$\eta$ is a Gaussian zero-average noise with correlation given by
$\left<\eta_t\eta_{t'}\right>=\Sigma_{tt'}$, where
$\Sigma=(\mathbbm{1}+G)^{-1}(\mathbbm{1}+C)(\mathbbm{1}+G^\top)^{-1}$ and
$C_{tt'}$ is the average spin autocorrelation between time $t$ and
$t'$.  This equation is not that of a representative agent, but is
representative of all the agents: one agent corresponds to a given
realization of the noise $\eta(t)$.

A further difficulty resides in extracting information from
Eq.\ \eqref{eq:eff_dyn}. In the asymmetric phase, one exploits the
existence of frozen agents, for which $Y(t)\propto t$ and assumes that
the stationary state correspond to time translation invariance
$X_{tt'}=X(t-t')$ for $X\in{G,C,D,\Sigma}$.  Thus, introducing the
notations $\tilde{X}=\lim_{t\to\infty}X/t$ and $\hat
X=\lim_{t\to\infty}\frac{1}{t}\sum_{t'\le t}X(t')$,
\begin{equation}
 \tilde Y=-\frac{\alpha}{1+\chi}s+\sqrt{\alpha}\hat{\eta},
\end{equation}
where $\hat{\eta}\sim\mathcal{N}(0,(1+\chi)^{-2}(1+Q_0)))$, and 
$Q_0=\hat{ C}$ and $\chi=\hat{tG(t)}$ correspond to the quantities
defined in the replica section; note that generating functions give a
precise mathematical definition of $\chi$. After some
lighter computations, one recovers all the equations of the replica
calculus; in addition, one also can discuss in greater rigor the
validity of simple expressions for $\sigma^2_0$.  The symmetric phase
still resists full analysis \cite{demartino2011nonergodic}, which
prompted the introduction of a further simplified MG, of the spherical
kind \cite{galla2003dynamics} (see Sec.~\ref{sec:mg_ext}).

The above sketch shows that the dynamics is ever present in the
equations; reasonable assumptions about some quantities may be made regarding their time
dependence or invariance, etc. This method allows one to control the 
approximations; it has confirmed the validity of the continuous time
equations and of the effective theory introduced in Sec.
\ref{sec:mg_continuoustime}.

The original MG gradually yielded to the power of generating functionals: first
batch MGs \cite{CoolenBatch}, then on-line MGs with random histories
\cite{CoolenOnline}, then on-line MGs with real histories
\cite{CoolenRealHistories}, which is a genuine mathematical {\em tour
  de force}; the case $S>2$ is treated in \textcite{CoolenS>2} and was later
simplified in \textcite{AdemarS>2}.

\subsection{Modifications and extensions}
\label{sec:mg_ext}
The MG is easily modifiable. Given its simplicity and many
assumptions, a large number of extensions have been devised and studied.  Two types
of motivation stand out: to determine in what respect the global
properties, e.g. herding, phase transition, etc., depend on the
strategy space and learning dynamics, and to remedy some shortcomings
of the original model.

\subsubsection{Payoffs}

The original MG has a binary payoff, which is both a curse for exact
mathematical methods and a blessing for ad-hoc combinatorial
methods. \textcite{MC01} show how to derive the quantity minimized by a 
MG with a generic payoff function $-a_iG(A)$; \textcite{C04} extends
this argument to explain why the location of the critical point is
independent on $G(A)$ as long as $G$ is odd \cite{SavitPayoff}, which
is confirmed in \textcite{papadopoulos2009theory}, who solved the
dynamics of the game for any payoff with generating functionals and
add that $G$ should also be increasing (see also Sec.~\ref{sec:mg_markets}).
The dynamics of the symmetric phase does
depends on the choice of payoff. For instance the game becomes
quasi-periodic only when a sign payoff is used
\cite{CM99,galla2005strategy}, and only for small enough $\alpha$
\cite{liaw2007three}.

From a mathematical point of view, the majority game can be considered
a MG with another payoff; $H$ is now maximized, in a way reminiscent
of Hopfield neural networks \cite{hopfield1982neural}, which makes
possible to use replicas \cite{kozlowski2003majGame} and generating
functionals \cite{papadopoulos2009theory}. Mixing minority and
majority players also yields to mathematical analysis
\cite{deMGM03,papadopoulos2009theory} and is discussed further in
Sec. \ref{sec:mg_markets}.

\subsubsection{Strategy distributions}

The thermodynamic limit only keeps the two first moments of the
strategy distribution $P(a_{i}^{\mu})$ (a consequence of the central
limit theorem). Its average must be rescaled, $\left\langle
a\right\rangle =\gamma/\sqrt{N}$, in order to avoid divergences; the
location of the critical point depends on both variance and average of
$P(a)$ \cite{CCMZ00,CMO03}.

The agents may draw their strategies in a correlated manner; for
instance, an agent may draw a first strategy at random as before, but
he chooses his second one so that $P(a_{i,1}^\mu=a_{i,2}^\mu)=c$, with
$c\in[0,1]$ \cite{MMM,garrahan2001correlated,galla2005strategy}.
 
Strategies may be used in a different way: \textcite{Oxf1} propose to
perform an inner product between a given strategy, considered a
vector, and a random vector living on the unity sphere; this model is
solved in \textcite{coolen2008inner}.

Sec.~\ref{sec:mg_markets} deals with strategies that also contain a
`zero' choice, i.e., the possibility to refrain from playing.

\subsubsection{Spherical Minority Games}

A special mention goes to spherical MGs \cite{galla2003dynamics} whose
dynamics is exactly and explicitly solvable in all phases while
keeping almost the same basic setup of the original MG; when using a generating function for the latter,
calculus is hindered by the non-linearity of
$s_i(t)=\textrm{sgn}\,Y_i(t)$: the boldest way to remove it is to set
$s_i=Y_i$. Because of Eq. \eqref{eq:a_omega_xi_si}, the agents may now
use any linear combination of their two strategies. Since $Y_i(t)$ may diverge, one adds the
spherical constraint $\sum_is_i^2=r^ 2N$.  This family of models also
undergoes phase transitions; its phase space $(\alpha,r)$ has a quite
complex structure.

Many extensions to the MG have been made spherical, thus, duly solved
\cite{galla2003dynamics,galla2005stationary,galla2005strategy,papadopoulos2008market,bladon2009spherical,demartino2011nonergodic}.

\subsubsection{Impact of used strategies  and Nash equilibrium}

The agents have several strategies to choose from and use only one at
a time.  A key point to understand why the agents fail to control the
fluctuations better in the symmetric phase is the difference of
expected payoff between the strategies that an agent does not use, and
the one that he plays. The discussion parallels that of the $P=1$ case
(see Sec.~\ref{sec:mg_P1} ): separating the contribution of trader
$i$ from $A$ in Eq.\ \eqref{eq:U_i_M>0} shows once again that self
impact results in payoffs that are biased positively towards the
strategies not currently in use and explains why all the agents are
not frozen in the original MG. Agents may experience difficulties in
estimating their exact impact; hence, \textcite{MCZe00} proposed to
modify Eq.\ \eqref{eq:U_i_M>0} to
\begin{equation}\label{eq:U_i_M>0_withimpact_eta}
U_{i,s}(t+1)=U_{i,s}(t)-a_{i,s}^{\mu(t)}(t)A(t)/N+\eta\delta_{s,s_i(t)}.
\end{equation}
Remarkably, the agents lose the ability to herd as soon as $\eta>0$:
$\sigma^2/N$ is discontinuous at $\eta=0$ in the symmetric phase; a
Nash equilibrium is reached for $\eta=1$ and all the agents are
frozen; there are exponentially (in $N$) many of them
\cite{demartino2001replicasymbreaking}; the one selected by the
dynamics depends on the initial conditions. The agents minimize
$H_\eta=(1-\eta)H+\eta\sigma^2$, which coincides with $\sigma^2$ when
$\eta=1$.  \textcite{MCZe00} noted that the difference between $H$ and
$\sigma^2$ is similar to an Onsager term in spin glasses
\cite{MPV}. When $H_\eta$ has no more a single minimum, the replica
calculus is more complex; one needs to use the so-called 1-step
replica symmetry breaking assumption (1-RSB) \cite{MPV}. 
\textcite{demartino2001replicasymbreaking} applies this method and reports 
the line at which $H_\eta$ ceases to have a single minimum, also
known as the de Ameilda-Thouless (AT) transition line
\cite{AT1978}. \textcite{AdemarHeimel} use generating functionals to solve the dynamics of 
Eq.~\eqref{eq:U_i_M>0_withimpact_eta} and discuss this transition from
a dynamical point of view by focusing on long-term memory and time
translation invariance. A simpler way to compute the AT line is given
in \textcite{MGbook}.

\subsubsection{Time scales and synchronization}

The original MG has two explicit intrinsic time scales, $P$ and
$\Gamma$, which are common to all the agents. There is a third one,
the time during which a payoff is kept in $y_i$, and is infinite by
default. Introducing a finite payoff memory  is easy if one discounts exponentially 
past payoffs, which amounts to writing
\begin{equation}
U_{i,s}(t+1)=U_{i,s}(t)\left(1-\frac{\lambda}{P}\right)-a_{i,s}^{\mu(t)}(t)\frac{A(t)}{N},
\end{equation}
where $\lambda\in[0,P]$ and the factor was chosen so as to introduce
$\lambda$ as a separate timescale; the typical payoff memory length
scales as $1/\lambda$ for small $\lambda$. This seemingly
inconspicuous alteration of the original dynamics changes very little
the dynamics of the asymmetric phase. It does however solve the
problem of non-ergodicity of the symmetric phase since initial score
valuations are gradually forgotten \cite{CDMP05}. Unfortunately, it
also has a great influence on analytical results, since an
infinitesimal $\lambda$ has so far prevented from obtaining any mathematical insight about the stationary state from
generating functionals: they still yield the exact effective agent
dynamics but nobody has found a way to extract information about the
stationary state because there are no more frozen agents
\cite{CDMP05,demartino2011nonergodic}. The spherical MG with payoff
discounting is of course exactly solvable with this method
\cite{bladon2009spherical,demartino2011nonergodic}. Replicas can be
applied in some cases: \textcite{marsili2001learning} study an MG with
impact and discounting; the quantity minimized by the dynamics is now
$\sigma^ 2+\frac{\lambda}{\Gamma}\sum_i
[\log(1-m_i^2)+2m_i\tanh^{-1}(m_i)]$; as the ratio $\lambda/\Gamma$
 between the memory and learning timescales increases,
the system undergoes a dynamical phase transition at
$\lambda/\Gamma\simeq 0.46$ between a frozen RSB phase and a phase in
which it never reaches a Nash equilibrium. Finally, the case $P=1$ is
easily solved with $\lambda>0$. For instance the critical learning
rate is $\Gamma_c=2-\lambda$: forgetting the past destabilizes the
dynamics as this decreases the effective over which past payoffs are averaged \cite{mosetti2006minority}.

There is converging evidence that human beings act at widely different
timescales in financial markets
\cite{LilloUtility,zhou2012strategies}.\footnote{The burstiness of
  human activity is another explanation to heavy-tailed activity of
  agents \cite{BarabasiActivity}.} In the context of the MG, they may
therefore differ in $P$, $\Gamma$ or
$\lambda$. \textcite{mosetti2006minority} split the populations in
subgroups that each have a different set of $\Gamma$ and/or $\lambda$,
for $P=1$: it turns out that it is advantageous to have a smaller
$\Gamma$ and a larger $\lambda$. In other words, to learn as little as possible
and to forget it as soon as possible, i.e., to behave as randomly as
possible. This makes senses, as a random behavior is a Nash equilibrium. Heterogeneity of $P$ is
studied e.g. in
\textcite{CZ97,CZ98,SavitPayoff,JohnsonEnhancedWinnings,MMM}.

Another way to implement heterogeneous time scales is to assume
introduce the possibility of not doing anything for some $\mu$, i.e.,
to generalize the probability distribution of $a_{i,s}^\mu$ to
$P(a)=f[\delta(a-1)/2+\delta(a+1)/2]+(1-f)\delta(a)$ \cite{Piai}; each
agent has an intrinsic frequency $f$ drawn from a known distribution;
agents that play frequently are less likely to be frozen. Replicas
\cite{Piai} and generating functionals \cite{demartino2003dynamics}
solve this extension.

Finally, the MG assumes perfect synchronization, which is a strong
assumption, but a useful one. Note that introducing frequencies as
discussed above is a cheap way to build partial synchronicity,
especially for small average value of
$f$. \textcite{mosetti2009structure} proposed a way to fully
desynchronize agent-based models; the maximally asynchronous MG keeps
its phase structure provided that the temporal structure of
interaction is not too noisy.

\subsubsection{Learning algorithm}

The common rationale of all learning schemes is that using them should
{\em a priori} improve the realized payoffs. Quite remarkably, the
literature on the MG has mainly considered variations of the theme
of the logit model, most often fixed  look-up tables, and simple
ad-hoc Markovian algorithms, ignoring the rest of the vast
reinforcement learning (RL) literature, which in passing goes against
the golden rule of learning: agents (including researchers) should
find the balance between learning and exploration
\cite{catteeuw2012heterogeneous}; see
\textcite{sutton1998reinforcement} for a superb review written at the
time of the introduction of the MG. In particular, $Q$-learning is
currently thought to mimic very well how human beings learn; 
see \textcite{montague2006imaging} for a review. It consists in exploiting optimally the
relationship between one's actions at time $t$ and the payoff at
future time $t+1$, conditionally on the states of the system at times
$t$ and $t+1$: the payoffs at time $t$ therefore also comprise some
future expected payoffs. The definition of states and actions are to
be chosen wisely by the authors: \textcite{MG-Qlearning} use look-up
tables $a_{i,s}^\mu$; the possible actions and state space are the
choice of strategy; this means that agent $i$ chooses $s_i(t)$ according
to a $Q$-learning rule; the resulting fluctuations are very similar to a
Nash equilibrium for look-up tables, though nobody has ever checked it accurately. 
\textcite{catteeuw2012heterogeneous} assume instead that the state is
$\mu(t)$ (real histories) and possible actions are $\{-1,+1\}$; they
also assume that the resource level $L(t)$ is a sinusoid and show that
$Q$-learning does very well in this
context. \textcite{catteeuw2009learning} considers a $P=1$ setting and
shows that $Q$-learning also converges to the Nash equilibrium $|A|=1$,
as do other very simple schemes from RL literature that are close to
the ad-hoc ones discussed in Sec. \ref{sec:mg_P1}; interestingly,
using $Q$-learning is a dominant strategy if the agents may select
their RL scheme by Darwinian evolution. No analytical results have so far been 
reported about these alternate RL schemes, although obtaining some
seems within reach.

Strategy exploration by the agents, i.e., letting the agents evolve
badly performing strategies, has been investigated in MG literature: a
look-up table is akin to a DNA piece of code, hence changing it is
akin to genetic mutations. \textcite{CZ98,SavitEv1} let the worst
performing agents replace their strategies, either at random, or by
cloning those of the best players;
\textcite{Sysi-Aho2003a,Sysi-Aho2003b,Sysi-Aho2003c,Sysi-Aho2004a} give 
to the agents the possibility of hybridization and genetic crossover of their own strategies; \textcite{CZ97,SavitEv2} allow
the agents to choose their memory length. In all these papers,
strategy exploration is beneficial to the agents and to the system as
a whole, and sometimes spectacularly so, 
see \textcite{Sysi-Aho2003a,Sysi-Aho2003b,Sysi-Aho2003c,Sysi-Aho2004a}.

In \textcite{Kinzel}, \textcite{Kinzel2}, \textcite{kinzel2002interacting}, the agents use
simple neural networks (perceptrons); the authors derive an analytical
expression for $\sigma^2$ as a function of the learning rate. They
also note that the neural networks have the peculiar task
of anti-learning, which tends to produce seemingly random outputs,
and discuss a possible application to cryptography.

%

\subsection{Minority Game and financial markets}
\label{sec:mg_markets}

The connection between financial markets and MGs is both strikingly
intuitive and deceptively hard to formalize clearly. At a high level,
it rests on the following observations:
\begin{enumerate}
\item Financial markets are competitive and their dynamics is similar
  to Darwinian evolution \cite{ZMEM,FarmerForce,lo2004adaptive}.
\item They are negative sum games, if only because of transaction
  costs.
\item They tend to have bursts of fluctuations (called volatility in
  this context).
\item They tend to be almost unpredictable because traders (human beings or algorithms)  tend to exploit and reduce price predictability.
\end{enumerate}

So far, the MG has all the ingredients needed to model the dynamics of
a model of price predictability dynamics, except a price
dynamics. Since $A$ is an excess demand or offer of something, assume
for the time being that it has some relationship with price evolution
(this point is discussed at length below). Then Fig.\ \ref{fig:s2HRS}
provides a very appealing scenario for the emergence of large
fluctuations in financial markets: predictable prices correspond to
mild fluctuations are bound to attract more traders who then
reduce $H$; once the signal-to-noise ratio becomes too small, the
agents herd on random fluctuations and produce large fluctuations. Large price fluctuations are therefore due
to too a small predictability. In other words, markets are stable as
long as they are predictable and become unstable if the
traders (i.e., money) are in play. \textcite{MarsiliInstabMarkets} also
  find the existence of a critical amount of invested capital that
  makes markets unstable in a very different model. This suggests in
turn that real markets should hover over  a critical point, which
explains periods of quiescence and periods of large fluctuations.

One of the shortcomings of the above picture is that $N$ is fixed in
the game, which implies some sort of adiabatic approximation. Adaptive
agents should be able to decide by themselves when they are willing to
play.\footnote{Players that decide not to take part to a game are
  called loners in game theory. Allowing for this possibility changes
  much the dynamics of even simple games, see e.g.,~\textcite{HauertLonerPRL}.} In a
financial market context, the agents must not only decide which is the
best strategy to play, but also if it is worth using it. In other words,
the agent's decision should rest not only on payoff differences
(e.g. $Y_i$), but also on the value of $U_{i,s}(t)$
\cite{SZ00,J99,J00}: this leads to the Grand Canonical MG (GCMG), in
which a reservoir of agents may or may not play  at a given time step
depending on whether one of their trading strategies is perceived as
profitable. This, in fact, mimics precisely for instance how
quantitative hedge funds behave. The learning algorithms that
  they apply are hopefully more sophisticated; for instance, some of
  them try to account for their impact on the price dynamics when
  backtesting a strategy.
  
  In the simplest version of the GCMG, the
agents have only one trading strategy $a_i^\mu$ and the possibility of
not playing; this is equivalent to having two strategies, one drawn at
random, and the zero strategy $a_0^\mu=0$ $\forall \mu$
\cite{CM03}. The score difference dynamics is
\begin{equation}
Y_i(t+1)=Y_i(t+1)\left(1-\frac{\lambda}{P}\right)-a_i^\mu(t)\frac{A(t)}{P}-\frac{\epsilon}{P}.
\end{equation}
The last term is a benchmark, i.e., the value attributed to not
playing. It is the sum of the interest rate and transaction costs, and
possibly of the willingness to play of a given agent. When
$\lambda=0$, $\epsilon=0$ does not make sense since an agent that 
comes in and then goes out of the game experiences a sure net loss. 
The typical timescale of the GCMGs is proportional to
$P/(\Gamma\epsilon\lambda)$.

Since the GCMG is a negative sum game, all the agents stop playing
after a while if the score memory length is large enough. In other
words, they need to feed on something. \textcite{MMM} introduce
additional agents with fixed behavior, called producers, who use the
markets for other purposes than speculation. The producers play a
negative sum game, but a less negative one thanks to the speculators,
which may play a positive game thanks to the producers. This defines a kind of market ecology
best described as a symbiosis \cite{MMM,ZMEM,CCMZ00}. One assumes that there are $N_s$ speculators and $N_p$ producers.

\begin{figure}
\includegraphics*[width=8.5cm]{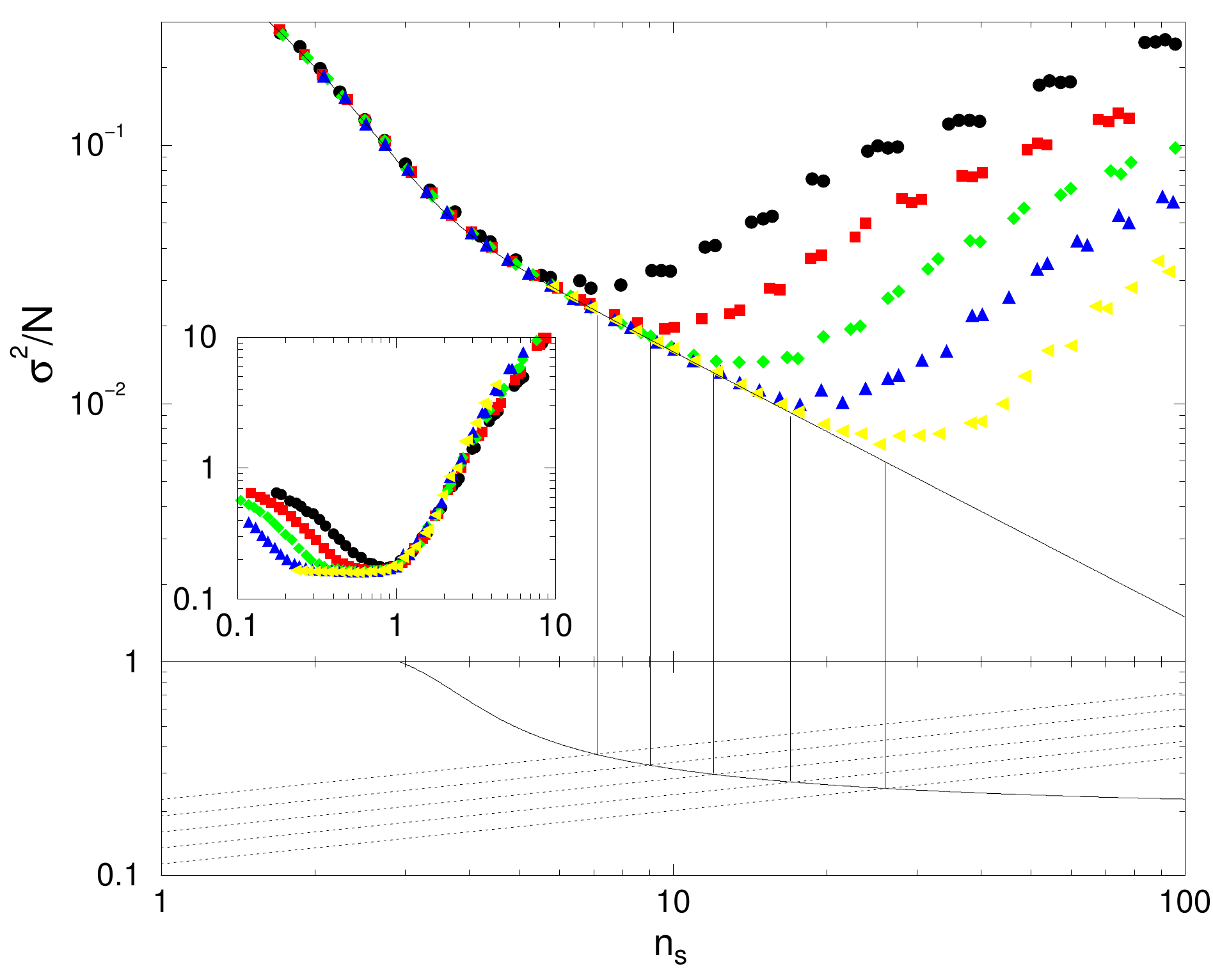}
\caption{\label{fig:gcmg_signalnoise} Top panel: Scaled fluctuations
  $\sigma^2/N$ versus $n_s=N_s/P$, where $N_s$ is the number of
  speculators and $N_p=P$ is the number of producers, shown for
  various system sizes $PN_s=1000$ (circles), $2000$ (squares),
$4000$ (diamonds), $8000$ (up triangles) and $16000$ (left triangles). 
Continuous line is exact solution for infinite systems.
Bottom panel:  LHS of Eq.~\eqref{eq:condvolclus} (continuous line) from 
the exact solution and $K/\sqrt{P}=K(n_s/L)^{1/4}$ (parallel dashed
lines) as a function of $n_s$ ($K\simeq 1.1132$ in this plot). 
The intersection defines $n_s^c(P)$. Inset: Collapse plot of 
$\sigma^2/N$ as a function of $n_s/n_s^c(P)$.
  From \textcite{CM03}.}
\end{figure}

For $\lambda=0$ and $\epsilon=0$, this model possesses a semi-line of
critical points $n_s=N_s/P>n_s^c(P)$: in other words, it is in a critical state as soon as there are
enough speculators in the reservoir. The signal-to-noise transition is
still present, which leads to anomalous fluctuations: using the method
described in Sec. \ref{sec:signal-noise}, one finds
\begin{equation}
\frac{H}{\sigma^2}+2\epsilon\sqrt{{H}{P}}\frac{P}{\sigma^2}+\epsilon\frac{P}{\sigma^2}\simeq
\frac{K}{\sqrt{P}},
\label{eq:condvolclus}
\end{equation}
which is confirmed in Fig.~\ref{fig:gcmg_signalnoise}; when
$\epsilon=0$, one recovers Eq.~\eqref{eq:sign-noise-MG}, thus
$\sigma^2/N$ behaves as in Fig.~\ref{fig:s2gamma}. When $\epsilon>0$,
the region of anomalous fluctuations shrinks as the system size
diverges; see \textcite{CM03,galla2006anomalous} for more details.
The $P=1$ version of the GCMG has additional instabilities compared to
a standard $P=1$ MG \cite{CAM08}.

Not only the distribution of $A$ becomes anomalous, but the strength
of fluctuations acquires a long memory. This is a feature generically found in MGs where agents can modulate their activity, either by re-investing a fraction of their gains, or by deciding to trade or not to trade. This result is even more generic:  \textcite{BouchaudGiardina} shows that {\em any} model in which the agents decide to trade or not depending on the sign of a random walk acquires automatically long memory in its activity and, by extension, to volatility. In the case of market-like MGs, whether to trade or not is based on the trading performance of a strategy. The agents that switch between being active and inactive have a strategy score that is very well approximated by a random walk.

The two possible actions $-1$ and $+1$ (and possibly $0$) may mean sell and
buy, respectively. In that case $A$ is an excess demand, which has an
impact on price evolution; \textcite{J00} use a linear price impact
function \cite{FarmerImpact,ContBouchaud}, $\log p(t+1)=\log
p(t)+A(t)$. This implies that $A$ is a price return.

But this raises the question of why the traders are rewarded to sell
when the majority buys, and reversely. There are two answers to
this. First, when an agent makes a transaction, being in the minority
yields on average a better transaction price \cite{MGbook}. Why should
an agent transact at every time step, then, unless he is a market
maker?\footnote{Market makers are special traders whose task is to
  propose transactions for buyers and sellers simultaneously, like
  {\em bureaux de change} for foreign exchange; they are thus most
  likely to transact very often.} \textcite{MarsiliMinMaj} argued that
the agents do not know which price they will obtain when they trade,
thus that they need to form expectations on their next transaction price:
the agents who believe that the price follows a mean-reverting process
play a minority game, while those who believe that prices changes are
persistent play a majority game. \textcite{deMGM03} therefore
introduced a model with minority and majority traders and give its
solution with replica and generating functions, later generalized in
\textcite{papadopoulos2009theory}.

There remains, however, an inconsistency: predictability is linked to
speculation, but the agents cannot really speculate, as their actions
are rewarded instantaneously.  This is why \textcite{BouchaudGiardina,dollargame} 
proposed to reward current actions with
respect to future outcomes, i.e., $a_i(t)A(t+1)$: this is a delayed
majority game whose peculiarity is that the agents active at time $t$ also play at time
$t+1$; it is known as the \$-game. The nature of this game depends on the sign of the
autocorrelation of $A(t)$: an anticorrelated $A$ causes an effective
minority game, and reversely; left alone, \$-game players tend to be
equivalent to majority players \cite{ferreira2005real,satinover2008cycles}. \textcite{BouchaudGiardina}define a more  realistic model and show that the price may be periodic (i.e., produce bubbles and crashes), stable, or intermittent (i.e. realistic) depending on the ratio $\Gamma/\lambda<~0.4$  and the contrarian/trend-following nature of the strategies.

And yet, modeling speculation must include at least two time steps:
its somehow counter-intuitive salient feature is that one possibly
makes money when waiting, that is, when doing
nothing. \textcite{ferreira2005real} stretched single time-step
look-up tables to their limit by assuming that agent $i$ whose action
was $a_i(t)$ at time $t$ must play $-a_i(t)$ at time $t+1$. If all
agents act synchronously, $A(t+1)=-A(t)$ and the \$-game becomes a
minority game. When some people act at even times and the others at
odd times, the nature of the market is more complex: in a mixed
population of minority/majority/\$-game players, the game tends to be
a minority game.

Modeling speculation requires to walk away from single time-step look-up tables. One
wishes however to keep a discrete number of states, which makes it
easy to define price predictability. \textcite{C05} still assumes that
the market states are $\mu(t)\in\{1,\cdots,P\}$, either random or
real; an agent can only recognize a small number of market states and may only
become active when $\mu(t)$ is one of them; he may invests between
pairs of patterns if he think it worthwhile. Accordingly, global price
predictability is now defined between all pairs of market
states. Price fluctuations, predictability and gains of speculators as a
function of the number of speculators are very similar to those of
GCMGs.

We believe therefore that the MG is the correct fundamental model to
study the dynamics of predictability, hence market ecology and their
influence on price fluctuations. Reversely, any correct model must
contain agents that learn, exploit and reduce predictability; it
therefore contains some kind of minority mechanism, be it explicit or
hidden. For instance, \textcite{hasanhodzic2011computational}
introduced independently a model of agents learning price
predictability associated to a given binary pattern and study how
information is removed; it is best described as a minority
game. Another attempt to define {\em ab initio} a model with producers 
and speculators in which the speculators remove
predictability \cite{patzelt2012unstable} is equivalent to the MG defined
in \textcite{galla2009minority}.

All MG models are able to reproduce some stylized facts of financial
markets; notably $P(A)\propto A^{-\gamma}$ and
$\left<|A(t)||A(t+\tau)|\right>-\left<|A|\right>^ 2\propto \tau^
{-\beta}$ allow their agents to modulate their investments
according to their success, as for instance GCMGs. In addition, evolving capitals and reinvestment have the
same effect and lead to power-law distributed $A$ at the critical
point for $S=2$ \cite{CCMZ00}, as well as for $S=1$
\cite{galla2009minority}. At this critical point, anomalous fluctuations are not finite size
effects. Even better, generating functionals solve the $S=1$ model; what
happens at the critical point awaits further investigations. 

Since the dynamics of market-like MGs is reasonably well-understood, one
may probe how it reacts to dynamical perturbations. The effect of 
Tobin-like taxes in a GCMG is akin to increasing the baseline
$\epsilon$; not only it reduces the occurrence of anomalous
fluctuations in the stationary state, but the dynamical decrease of
anomalous fluctuations in reaction to a sudden increase of $\epsilon$
is very fast \cite{bianconi2009tobin}. On the other hand,
\textcite{papadopoulos2008market} introduced a constant or periodic
perturbation to $A(t)$ in a spherical MG; the effect of such a 
perturbation is counter-intuitive:  $A(t)$ may lock-in in
phase with the perturbation, which increases fluctuations.  A third
work investigated the effect of a deterministic perturbation that lasts for
a given amount of time in the non-spherical GCMG; this corresponds
to a sometimes long series of transactions of the same kind (e.g., buy) known as
meta-orders; see \textcite{BouchaudFarmerLillo} for a review. Using
linear response theory and results from the exact solution of the
game, \textcite{barato2011impact} computed the temporal shape of the
impact on the price to expect from such transactions.

There is yet another way of understanding the relationship between the
MG and financial markets \cite{C05}: on a abstract level, $A(t)=0$ corresponds to
perfect coordination, as it is an equilibrium between two opposite
actions. These two actions may be to exploit or not  gain
opportunities, labeled by $\mu$. If too few
traders exploit it, more people should be tempted to take this
money-making opportunity; if there are too many doing so, the realized trading
gain is negative. In this sense, the MG is connected to trading, since market
participants use a trading strategy that exploits a set of gain opportunities that seems profitable only if under-exploited. In this cas, a minority mechanism is found because people try to learn an implicit ressource level.

\textcite{JohnsonLargeChanges} apply the GCMG to the prediction of
real market prices by reverse-engineering their time-series. They propose to find 
the specific realization of the GCMG that reproduces
some price changes over a given period most accurately and to run it a few
time-steps in advance. Large cumulative price changes produced
by the game are reportedly easily predictable. According to
\textcite{SornettePocket}, these pockets of predictability come from
the fact that sometimes many agents will take the same decision $k$
time-steps in advance, irrespective of what happens between now and
then. Some more statistical results about the prediction performance of 
minority/majority/\$-games are reported in
\textcite{wiesinger2010reverse}. A few other papers use modified MGs
in the same quest \cite{krause2009evaluating,ma2010minority}. The same
principle was used to predict when to propose discounts on the price
of ketchup \cite{MGketchup}. As emphasized by \textcite{J00}, the
whole point of using non-linear adaptive agent-based models is to
profit from strong constraints on future dynamics to predict large
price changes; this goes way beyond the persistence of statistical
biases $\left<A|\mu\right>$ for some $\mu$'s.

Making the connection with more traditional mathematical finance,
\textcite{ortisi2012minority} assumed that the price dynamics of a financial asset 
is given by the continuous-time dynamics of the vanilla MG and GCMG, computed analytical
expressions of the price of options,
and proposed a method to calibrate the MG
price process to real market prices.

Finally, all the previous papers focus on a single asset, but most
practitioners wish to understand the origin and dynamics of price
change cross-correlations. \textcite{bianconi2008multiassetsMG} gives
to the traders the opportunity to choose in which game, i.e., assets,
they wish to take part, both for the original MG and for the GCMG; more
phase transitions are found depending on how much predictability is
present in either asset; generating functionals solve the lot;
\textcite{AdemarS>2} extended to this calculus to more generic ways of
choosing between many assets .

\subsection{Multichoice Minority Games}

Extending the MG to more than two choices seems easy: it is enough to
say that $a_i^\mu$ may take $R>2$
values. \textcite{Kinzel3}, \textcite{chow2003multiplechoiceMG}, \textcite{quan2004evolutionarymultichoiceMG}
consider $R\ge 3$ and reward agents that select the least crowded
choice; \textcite{dhulst1999threesidedMG} introduce cyclical trading
between three alternatives.

There may also be $R$ types of finite resources, e.g., bars:
\textcite{savit2003finiteresources,savit2005generalallocgames} assume
that $N$ agents choose between $R=2$ types of resources, each of them
able to accommodate $L$ agents. This situation arises in CPU task
scheduling.  \textcite{shafique2011minorityCPU} take the reverse point of view: the agents may be groups of CPU cores
competing for tasks to execute. A more
complex structure underlies multi-assets models: the agents first
choose in which asset to invest, and then play a minority game with
the other agents having made the same asset choice
\cite{bianconi2008multiassetsMG}.

Whereas these studies assume that $R$ is fixed while $N$ may be
arbitrarily large, many real-life situations ask for $R$ to scale linearly with
$N$: this is the assumption of Secs. \ref{sec:kpr} and
\ref{sec:bipartite}.

\subsection{Minority mechanism: when?}

The definition of the MG is quite specific and seems to restrict a
priori in a rather severe way its relevance. We wish to suggest a more
optimistic point of view. There are  universal relationships between
fluctuations and learning in MGs. Therefore, should a minority mechanism 
be detected in a more generic model, one can expect to understand which part of its global properties come from the minority mechanism. This requires to understand what a minority mechanism really is and where it may hide.

EFBP does contain one, since it is a MG with a generic resource level $L$
\cite{CMO03,JohnsonAsym}. This resource level may depend on time \cite{galstyan2003resource}; the
resulting fluctuations will come both from the transient convergence
to a new $L$ and from fluctuations around a learned $L(t)$, which are
of the MG type. This view indicates that a minority mechanism arises
when a population self-organizes collectively around an explicit
resource level. Self-consistent resource levels sometimes contain minority mechanisms:
 \textcite{C04} considers one population of producers pooling their contributions  $A=\sum_ia_i$ and one population of buyers grouping their monetary offers $B=\sum_k b_k$ for the production on offer. Producers should decrease their output if $A>B$ and buyers should do so when $A<B$. This suggests a payoff $-a_i(A-B)$ to
producer $i$ and $-b_i(B-A)$ to buyer $i$, hence, that $B$ is the resource
level for the producers and $A$ is the resource level for the buyers, both time-dependent and self-consistently determined. The stationary state of this model is equivalent to the EFBP and is exactly solvable.  

In conclusion, one may expect a minority mechanism, hence, MG-like
  phenomenology in a situation where a population self-organizes
  collectively around an explicit or implicit resource level that it
  may contribute to determine. 

Let us now mention a few papers which seem a priori to have little to
do with the MG, but that do contain a minority mechanism. This helps
understanding their phenomenology, but cannot describe quantitatively their behavior, 
which may be much more complex and richer.

\textcite{cherkashin2009reality} introduced a game with two choices (in
its simplest form) whose mechanism is stochastic: the probability that
choice $+1$ is the right one is $P(A/N)$ (in the notations of this
paper), with $P(0)=1/2$: this introduces {\em mechanism noise},
but the average nature of the game is easy to determine: indeed, the
expected payoff of agent $i$ is
$\left<u_i\right>=2a_i(t)[P(A(t)/N)-1/2]$; introducing $G(A)=2P(A)-1$,
and expanding $G$, one finds that
$u_i(t)=a_i(t)[G'(0)A/N+O(A^3/N^3)]$. Clearly, minority games appear
when $P'(0)<0$, which is what the authors call self-defeating
games. One thus rightly expects that learning reduces
predictability $|A|$ in the latter case and increases it in the other
case.  A closely related extension of learning in MGs is {\em decision
  noise}, which causes the agents to invert their decision with
some probability; see e.g. \textcite{CoolenBatch}.

\textcite{Berg} introduced a model in which agents receive partial information about the real market state $\mu(t)$: they each are given their own projection function
of $\mu=1,\cdots,P$ onto two possible states $0$ and $1$, denoted by
$f_i(\mu)$ and randomly drawn at the beginning of the game. The resource level is
$R^\mu$, the market return when the global
state is $\mu$. Each agent effectively computes two averaged resource
levels $\overline{R^\mu|(f_i(\mu)=0)}$ and $\overline{R^\mu|(f_i(\mu)=1)}$
and finds out  how much to invest conditionally on $f_i(\mu)=0$ or $1$. 
Remarkably, when there are
enough agents, the prices converge to $R^
\mu$. The phenomenology of such models is different, but similar to that of the
MGs. Accordingly, some parts of the exact solution, both with replicas \cite{Berg}
and generating functionals \cite{demartino2005asymmetricinformation},
are quite similar to those for $H$ in the MG.

\section{The Kolkata Paise Restaurant Problem}
\label{sec:kpr}

In Kolkata (formerly Calcutta), there used to be cheap and fixed
rate ``Paise'' \footnote{\textit{Paise}, the smallest monetary unit in
  Indian currency, essentially synonymous with anything very cheap.}
restaurants which were popular among daily workers.  During lunch
hours, the workers used to walk, to save the transport costs, to one
of these restaurants. They would miss lunch if they got to a
restaurant where there were too many customers, since walking to the
next restaurant would mean failing to resume work on time!

\textcite{kpr-physica}, \textcite{mathematica}, \textcite{kpr-proc}, \textcite{kpr-njp} proposed the Kolkata Paise Restaurant (KPR) problem. It is 
a repeated game between $N$ prospective
customers that choose from $R$ restaurants each day
simultaneously (in parallel).  $N$ and $R$ are both assumed to be large and proportional to each other.
Assuming that each restaurant charges the same price for a meal eliminates budget constraints for the agents.

An agent can only visit one restaurant each day, and every restaurant
has the capacity to serve food to one customer per lunch
(generalization to a larger value is trivial).  When many agents
arrive at one restaurant, only one of them, randomly chosen, will be served. The main measure of global efficiency is the utilization fraction $f$ defined as the average fraction of restaurants visited by at least one customer on a given day; following the notations of the literature on this topic, we denote by $\bar{f}$ its average in the steady state.  

Two main points have been addressed: how high  efficiency can depend on the learning algorithm and the relative number of customers per restaurant, denoted by $g=N/R$, and at what speed it is reached. 

Additional complications may arise if the restaurants have different
ranks.  This situation is found for instance if 
there are hospitals (and beds) in every town but the local
patients understandably prefer to go to hospitals of better rank
elsewhere, thereby competing with the local patients of those
hospitals.  Unavailability of treatment in time may be considered as a
lack of service for those people and consequently as social wastage
of service by the unvisited hospitals. This is very similar to the stable marriage problem briefly reviewed in Sec.\ \ref{sec:marriage}.

The most efficient solution to the KPR problem is dictatorship: each customer 
is assigned a restaurant and must eat there. If the restaurants have a ranking,  the customers must take turns and 
sample each of them in a periodic way, which is socially fair. Dictatorship however requires a 
level of synchronization and communication that is quite unrealistic. This is why the agents need to learn to try to have meals at the best ranked restaurants. Inevitably, these strategies take
 more time to converge to a less efficient state than under dictatorship, but some stochastic strategies are  better than others, reaching fairly efficient states in $\mathcal{O}(\log N)$ meals.

The natural benchmark of efficiency is obtained when all the agents choose at random and with uniform probabilities where to have lunch. The probability that a given restaurant is chosen
by $m$ agents is  binomial $\Delta(m) = \binom{N}{m} p^m
(1-p)^{N-m}$ with $p=\frac{1}{R}$.  In the limit $N,
R\rightarrow\infty$, keeping and $g=N/R$ finite, it becomes a Poisson
distribution $\Delta(m) = \frac{g^m}{m!} \exp({-g})$.  Clearly,
the fraction of restaurants not chosen by any agent is $\Delta(m=0) =
\exp(-g)$. Finally the average fraction of restaurants
occupied on any day is $\bar{f}= 1- \exp(- g)$; in particular $\bar{f} \simeq
0.63$ for $g=1$ \cite{kpr-physica}.

\subsection{As many restaurants as customers, $g=1$}

It is worth first to investigate the dynamics of the case where there are as many restaurants as customers ($R=N$, i.e., $g=1$).

\subsubsection{Stochastic rank dependent strategies}
Let the restaurants have a well-defined ranking (agreed by everyone)
depending upon quality of food, services, etc.,  although the price of a
meal is the same for all restaurants.  Thus, all agents will try to get
food from best rank restaurants.  But since a restaurant can serve
only one customer, it means that many of the agents in crowded
restaurants will remain unsatisfied.  Now, assume that any agent
chooses the restaurant of rank $k$ with probability $p_k(t)=k^\zeta/\sum_k
k^\zeta$, where $\zeta$ is a real number.

It can be easily shown that the probability that a $k$-th ranked
restaurant is chosen by no one is $ \Delta_k(m=0)=\exp\left(-{k^\zeta
  \left(\zeta+1\right)\over N^\zeta}\right)$, and hence the average
fraction of agents enjoying their meal there is given by $\bar f_k=1-
\Delta_k\left(m=0\right)$.  The limiting case $\zeta=0$ is equivalent
to random choice, hence, $\bar f_k=1-e^{-1}$, giving $\bar f=\sum_k \bar f_k/N
\simeq 0.63$. When $\zeta=1$, $\bar f_k=1-e^{-2k/N}$ giving $\bar
f=\sum_k \bar f_k/N \simeq 0.57$~\cite{kpr-njp}.

\subsubsection{Crowd-avoiding behaviors}
\paragraph*{Fully crowd-avoiding case:}
 Assume that all the agents avoid the restaurants too crowded at the previous lunch. In the stationary state, there are hence
 $N(1-\bar f)$ available restaurants for the next lunch, assumed to be randomly chosen by all the $N$
 agents. Using the fact that $N/R=1/{(1-\bar f)}$, one can write the recursion
 $(1-\bar {f})\left[1-{\rm exp}\left(-\frac{1}{1-\bar
     {f}}\right)\right]=\bar {f}$, which gives $\bar f\simeq 0.46$ and is worse than random choice, but typical of herding, as in the MG \cite{kpr-njp,kpr-proc}.

\paragraph*{Stochastic crowd-avoiding case:} 
Higher efficiency is reached when the agents do not systematically avoid previously crowded restaurants.
 Suppose that restaurant $i$ was chosen by $n_i(t-1)$ customers at time $t-1$.  The probability to return to restaurant $i$, denoted by $p_i$, should be such that $p_i(1)=1$ and be a decreasing function of $n_i(t-1)$. The simplest choice is $p_k=1/n_k(t-1)$, which yields $\bar{f}\approx 0.79$
\cite{kpr-njp}, a sizable improvement over the fully crowd-avoiding case. The time needed to reach the steady state is
$\sim \log N$.

A simple argument gives an approximate estimate of $\bar{f}$ in this case: 
Let $a_i$ be the fraction of restaurants visited by $i$
agents and assume that the fraction of restaurants where $3$ or more
customers arrive on any day are negligible, i.e., $a_i=0$ for $i\geq 3$.
One can write $a_1+2a_2 = 1$ and $a_0+a_1+a_2 = 1$, giving $a_0=a_2$.
According to this strategy, every agent who visited any restaurant
alone on a given day will surely return there next day ($p=1$), but if
any restaurant was visited by two then the agents will return there
with probability $p=1/2$ the next day. In this process every time
$(a_2/4-a_2^2/4)$ fraction of restaurants will be vacant from the
restaurants previously visited by $2$ agents.  Similarly $a_0a_2$
fraction of restaurants will be visited by one agent from the
restaurants previously visited by two. Therefore, one can easily write
$a_0-a_0a_2+ \frac{a_2}{4}-\frac{a_2^2}{4}=a_0$.  The above three
equations gives $a_0=0.2$, $a_1=0.6$ and $a_2=0.2$.  So the average
fraction of restaurants visited by at least one agent is
$a_1+a_2=0.8$, close to that obtained in the numerical simulations.

The above behavior can be generalized by letting $p_i=1/n_{i}^{\xi}(t-1)$
 where $\xi$ is real and positive. The case discussed above is recovered with $\xi=1$, while $\xi =0$ prevents any change of restaurant by the agents.  Numerical simulations show that both the utilization
fraction and the time to reach the steady state increases when $\xi$ decreases. The best utilization fraction is obtained when $\xi \to 0$, while it decreases to $\bar f \simeq
0.676$ when  $\xi \to \infty$ \cite{Ghosh2013}.

\subsection{Plethora of restaurants ($g< 1$) and phase transitions}
\label{subsec:phase}
\textcite{kpr-pre12} considered a KPR problem with more restaurants than customers.
 Let $n_i$ agents visit restaurant $i$ on a particular
day.  Each restaurant can serve only one agent in a day.  One
of the $n_i$ agents is chosen randomly and served while the remaining
($n_i-1$) agents do not get any lunch for that day.  Two variants of the return probability have been
investigated. In model A, the probability to return to restaurant $i$ 
 for each of the $n_i$ agents is $p_i=1/n_i$, as before.  In model B,  $p_i(1)=1$, but customers of crowded restaurants will return at the same place with uniform probability ($p(n_i)=p<1$, $n_i>1$);  $p$ can be thought of as the `patience' of an agent. 
Both models undergo a phase transition at some
$g=g_c$ below which all the agents will be satisfied; when $g>g_c$,
some are not.  Accordingly, the order parameter is  the
density of restaurants having more than one customer in the steady state, and denoted by $\rho_a$. 

Both mean-field and finite dimensional lattice versions ($1$D and $2$D) of these model have been studied. They turn out to belong in the same universality class as fixed energy sandpiles \cite{Vespignani}: 
\begin{enumerate}

\item Model A has $g_c=1$ for 1D, but $g_c < 1$ for higher dimensions.  The
critical exponent values for the phase transition in finite dimensions
are in good agreement with those of stochastic fixed-energy
sandpile~\cite{Vespignani,Manna:1991,Lubeck:2004}. 
\item  Model B has two
parameters, the density $g$ and the patience $p$ of agents.  The phase
diagram has been numerically investigated.  In the mean field case,
the phase boundary is linear: $g_c =\frac{1}{2}(1+p)$.  For $p=0$,
$g_c=1/2$, as in fixed energy sandpiles~\cite{Vespignani}.  The phase
boundary in ($g,p$) is nonlinear for 1D and 2D.  The critical
exponents are the same along the phase boundary and match with
those of model A.
\end{enumerate}

Note that when $g = 1$, the system is far from its critical value $g_c$ and the relaxation
time $\tau$ does not show any $L = N^{1/d}$ dependence.  As long as
$g\le g_c$, absorbing frozen configurations are present, and whether
they are accessible or not depends on the underlying dynamics. 
A critical point $g_c$ was found to exist above which the agents are
unable to find frozen configurations. 
The role of $p$ in model B is as follows:  when $g>g_c$, the agents fail to reach satisfactory
configurations -- they are moving only if they
are unsatisfied ($p=0$). More patient agents, i.e., those with a larger $p$, are linked to larger 
 values of the critical point $g_c$ and thus
smaller time to reach saturation (faster-is-slower effect).

\section{Generic bipartite allocation problems}
\label{sec:bipartite}
The KPR model is but an example of the more generic bipartite
allocation problems where one type of items must be associated to
another type of items; all the agents are homogeneous and the
restaurants are as well. More generically, the agents may be
heterogeneous and the items to choose from may also be heterogeneous.

\subsection{Spatial allocation problems}

An example of allocation of spatially aligned resources is the parking
lot problem \cite{Hanaki2011}. This model is a stylized version of an everyday experience of
  the difficulty in finding a parking spot for a car, as motivated in
  the work from the example of the small streets leading to
  \textit{Vieille Charit\'{e}} in Marseille, and could be the case for
  any other street.  One has to walk the rest of the way
upon finding a spot, costing both time and effort.  It is most enjoyable to
park close to the office, if luck permits, while the unlucky ones park
far away, regardless of the instant when they arrive.

Formally, in a hypothetical city, there is a unique one-way street allowing 
parking and leads to the center.  
$R$ parking spots available along the street are indexed by their distance from the center
$s \in \{1,2, \ldots, R \}$.  In the beginning of each period $t \in
\{0,1,2, \ldots \}$, each of $N \ge R$  agents can be either
outside or in the city, and parked. If agent $i$ is outside,
he drives into the city with probability $p_e$ and looks for a
vacant spot. His strategy  is first to fix his comfort distance  $s_i(t)\in\{1,\cdots,R\}$ and to take the first empty spot $k_i(t)<s_i(t)$. The closer to the center, the more rewarding a parking space is, thus being able to park at spot $k$ gives a payoff $\pi(k)$,  a decreasing function with boundary conditions $\pi(1)=1$ and $\pi(R)=0$.  Not finding a parking space yields a penalty of $-B<0$ and is denoted by $k_i(t)=0$. At then end of each round, each agent leaves the city
with probability $p_l$.  

Learning takes place in a familiar way.  The agents update
the score they associate with strategy $s$,   $U_{i,s}^i(t)$ and use the logit rule for their next choice
\begin{equation}
P[s_i(t) = s]=\frac{e^{\Gamma U_{i,s}(t)}}{\sum_{s'=1}^R e^{\Gamma
  U_{i,s'}(t)}},
\end{equation}
Initial scores are equal to the average payoff of parking along the street, $U_{i,s}(t=0) =
(1/R) \sum_{s'=1}^R \pi(s')$ $\forall i,s$.  The attraction of
strategy $s$ evolves, if the agent was trying to find a suitable vacant spot at time $t$, as
\begin{eqnarray}
U_{i,s}(t+1)&=& U_{i,s}(t)(1-\lambda) \\\nonumber
&& + \lambda (\pi[k_i(t)]\Theta[k_i(t)-s_i(t)+1/2] \\\nonumber
&& -B\Theta [-1/2+k_i(t)]).
\end{eqnarray}
The case  $N=R=2$ could be exactly solved~\cite{Hanaki2011}. One relevant question is whether the 
population learns to segregate itself into lucky people that are almost systematically able to park close to the center and unlucky ones. In order words, one should investigate if the steady state corresponds to a symmetric or asymmetric state. It turns out that the answer depends on both $\Gamma$ and $B$. If  $\Gamma > 4$, the
symmetric state is stable either for small cost $B$ where both players
choose strategy $1$, or for very large $B$, where both players choose
strategy $2$.  This implies that for low $B$, the street parking spots
are not fully utilized, while for high $B$, they are fully utilized
either because the two players choose different strategies, or
otherwise both choose to park further away. The large $N$ case is dealt with numerical simulations. The
small $B$ symmetric outcomes disappears. There is instead 
a large degree of heterogeneity even at $B=0$, which becomes small
for large $B$, and the agents learn to choose strategies with high
$s$, i.e, park further away. Thus, for high $B$, agents are rarely lucky.

\subsection{Stable marriage}
\label{sec:marriage}
A famous a priori symmetric example is the stable marriage problem
proposed by \textcite{gale1962college}, \textcite{knuth1997stable} where an equal number of men
and women must be matched according to their private preference list
that ranks all members of the opposite kind. This clearly
  also (and rather) applies to other matching problems such as
  colleges and students, hospital and doctors, etc. Configurations
are deemed stable if no pair of entities have an incentive to swap
their associated partner/item.  Mean-field analysis of the algorithm proposed
by Gale and Shapley showed that the dissatisfaction, or equivalently,
energy, of both sides obeys a simple scaling law
\cite{omero1997marriage}. Later extensions broke the a priori symmetry
in preferences list by introducing correlations, partial lists, match
makers and bachelors
\cite{caldarelli2001beauty,zhang2001happier,laureti2003matching}.

\subsection{Recommendation systems}

The assumption of complete preference lists is indeed very strong and
most real-life situations do not allow for such a luxury: for
instance, merchant websites such as Amazon or Netflix try to guess
what each of their clients might like and recommend them some other
items based on their previous purchases and browsing history. In this case,
preference lists are necessarily incomplete, both for the merchant and for the customers, and one must use
recommendation systems to guess their missing parts. It is beyond our
aim to review this sizeable research area; see
\textcite{lu2012recommender} for a recent review. We will briefly review
some concepts of physics applied to this family of problems. 

 One can think of
links between customers and items as bipartite networks. As the
fraction of known opinions increases, that is, as the network
connectivity increases, prediction undergoes two percolation-type
phase transitions, and perfect prediction may be possible in a simple
case, as shown in \textcite{maslov2001extracting}.  Another idea
consists in letting probabilities of liking items diffuse like heat on a 
suitably defined network \cite{zhang2007heat,zhou2010solving}.  
A recent development emphasizes the role of
competition for items: most recommendation systems assume that the
items to be chosen from are in infinite supply (or equivalently whose
duplication is essentially free, e.g., an mp3 file) and their
recommendations suffer from a lack of diversity: they tend to
recommend the same few items to everybody. But competition is a
fundamental driver of choice in resource allocation
problems. \textcite{gualdi2013crowd} show that assuming infinite
supply leads to a kind of bosonic condensation of recommendations
linked to a degeneracy of the cost function. Competition is introduced
by assuming that the potential gain of consuming some item decreases
when more people use them; this is enough to lift degeneracies, fight
biases and improve much the precision of recommendations even if the
items are infinitely replicable. A single parameter allows for
interpolating between the pure bosonic and fermionic cases.

\section{Discussion and outlook}

In this review we applied a variety of concepts 
from physics  to competitive resource allocation. 
We emphasize that our review limits to models which are characterized
by strong heterogeneity and non-equilibrium dynamics amongst adaptive agents, 
with minimal or no interaction.
Although a population of traders with minimal interaction in a certain class of
problems  can be treated as a gas of atoms~\cite{Chakrabarti2013},
here we stress  that  our systems are more complex and  
more sophisticated methods of statistical physics are required
to handle them.
Methods from statistical mechanics are well suited to the study of competition for limited resources as shown by the variety of models reviewed above. This is because most of them are of mean-field nature and involve many agents and/or many resources. The ability of generating functionals to solve the dynamics of minority games with real histories is quite impressive.

The basic MG model, together with some of its many extensions, is remarkably well understood. There are nevertheless areas in which better understanding or exact solutions would be much welcome:
\begin{enumerate}
\item
A low-hanging fruit is about MGs where each agent's weigth in the game is modulated according to his past success. They have been only superficially studied. Nevertheless, the coincidence between power-law distributed agent weights and the critical point is only alluded to in \textcite{galla2009minority}, who even provide the generating functional solution. This point is worth a detailed investigation in our opinion.

\item The non-ergodicity of the
symmetric phase still resists a full solution; it is cured by the
introduction of a finite score memory $1/\lambda$ that is enough to anihiliate the prowess of generating functionals because extracting useful information about the stationary state rests on the assumption that some agents are frozen, i.e., only use a single strategy. We believe that the problem lies in the definition of a frozen agent: instead of assuming that this means that he sticks forever to his favourite strategy, one should instead regard him as frozen as long as the typical duration of strategy exclusive use is larger than $1/\lambda$, which is the typical time scale of score memory in this kind of games.

\item
Investigating more learning rules in the MG is certainly worth trying. Indeed, the statistical mechanics community working on agent-based models has stayed too close to the logit model and has insufficiently explored reinforcement learning literature. Solving a model where heterogeneous agents use $Q$-learning and regret theory stands out as a particularly tempting problem since this combination is believed to best describe how human beings learn \cite{lohrenz2007neural}. This combination of payoffs and learning scheme seems to be well within reach of our tools.

\end{enumerate}

Reverse-engineering financial markets with agent-based models such as the MG is indubitably seducing. Although this idea was mainly proposed to predict future time series, one must realistically first ask oneself if the instance of the MG obtained by reverse-engineering is a faithful representation of the reality. When the time series to approximate is produced by a MG, this works well \cite{JohnsonLargeChanges} and yields a model that can predict future of MGs \cite{JohnsonLargeChanges,JohnsonPrediction}. But in the case of financial markets, comparing the strategies of the resulting best model with those of real traders has never been performed, for good reasons: this requires to have data about real traders. Fortunately, this is becoming ever more common \cite{MC10}. 

However, the models used to fit MGs to financial time series must be modified in two ways. First, assuming that all the traders react to each state of the market $\mu$ once they are in the market is probably sub-optimal. One should allow the strategies of the agents not to play for certain $\mu$ along the lines of \citep{Piai}.  In addition, most market-like models make the pathological assumption that an artificial traders cannot invert strategies are perform consistently badly, i.e., use $-a^\mu$ for all $\mu$'s if that is preferable. In an market-like MG with 1 strategy $a_i$ per agent and the possibility of not playing, this amounts to add $-a_i$ to this strategy set. When trying to reverse-engineer market price time-series, this accelerates the whole process by a factor of two. Talking about acceleration, the payoff updates of a GCMG with one strategy  (and its inverse) can be written using vectors, which makes this model an ideal candidate for easy GPU acceleration.

Physicists are sometimes
asked by somewhat skeptical and possibly conservative economists,
statisticians, etc. what may remain from their incursions in the
respective domains. The two key answers are  collective phenomena
caused by the interaction of many agents, and the mathematical methods
to understand and solve non-linear and disordered
systems.   

Other fields have also devised methods that exploit the mathematical simplifications 
provided by the large-system limit, i.e., by the central limit theorem. Economists have noticed that in suitably
linear models with logit learning, an effective dynamical equation may
be derived by averaging over population heterogeneity
\cite{brock2001largetypelimit}. Mathematicians propose the so-called
mean-field games theory that constructs the Hamilton-Bellman-Jacobi
equations equivalent to the dynamics of interacting optimizing
sub-populations over a finite time horizon
\cite{lasry2007meanfieldgames}. These two methods are more restrictive than
generating functionals which can solve (in principle) the dynamics of
any mean-field adaptive agent-based model with complex heterogeneity
and complex learning schemes, without any explicitly optimized
function. They are however under-used, under-understood and
under-publicized beyond the small community of people knowledgeable
about both statistical mechanics and complex systems. This may come
from their apparent mathematical complexity, whereas they involve
little more than very long expressions involving Gaussian integrations
and very many auxiliary variables. Part of the work that lies ahead
is to propagate these mathematical methods beyond our field.

\section*{Acknowledgments}

We would like to thank all our collaborators: 
F. Abergel, 
A. C. Barato, 
M. Bardoscia, 
J. Berg, 
G. Bianconi, 
S. Biswas,  
M. Blattner,
I. P. Castillo,
A. S. Chakrabarti, 
S. R. Chakravarty, 
A. Chessa, 
D. Dhar, 
M. Dzierzawa, 
A. de Martino,
D. De Martino,
F. F. Ferreira,
T. Galla, 
A. Ghosh, 
S. Gualdi,
A. Hanaki, 
C. Heung, 
K. Kaski,
A. Kirman, 
P. Kozlowski,
Z. Kuscsik,
P. Laureti,
J.-G. Liu,
L. L\"u, 
S. Maslov,  
I. Mastromatteo, 
C. Matzke, 
M. Medo, 
M. Mitra,
D. Morton de Lachapelle,
G. Mosetti, 
R. Mulet, 
I. Muni Toke, 
T. Naskar,
M.-J. Om\'ero, 
G. Ottino, 
M. Patriarca, 
M. Piai, 
P. Pin,
G. Raffaeli, 
F. Ricci-Tersenghi,
A. Rustichini, 
V. Sasidevan, 
S. Sinha,
F. Slanina, 
S. Solomon, 
M. Sysi-Aho,
J. R. Wakeling,
C. H. Yeung,
Y.-K. Yu,
R. Zecchina,
Z.-K. Zhang,
T. Zhou.
We are also grateful to J.-P. Bouchaud and J. D. Farmer for stimulating discussions over the years.

\bibliographystyle{apsrmp4-1} \bibliography{biblio}

\end{document}